\documentclass[preprint,showpacs,preprintnumbers,amsmath,amssymb,nofootinbib]{revtex4-1}

\usepackage{graphicx}
\usepackage{epstopdf}

\usepackage{dcolumn,subfigure}
\usepackage{bm}
\usepackage{color}

\usepackage{feynmf,appendix}
\usepackage{slashed}
\usepackage{enumerate}

\usepackage[scientific-notation=true]{siunitx} 

\usepackage[utf8]{inputenc}

\usepackage[colorlinks=true,urlcolor=blue,anchorcolor=blue,citecolor=blue,filecolor=blue,linkcolor=blue,menucolor=blue,linktocpage=true]{hyperref} 

\usepackage{appendix}
\usepackage{amsmath,amsthm,amssymb,mathrsfs}
\usepackage{bbold}
\usepackage{bbm}

\usepackage{multirow}
\usepackage{makecell}
\usepackage{verbatim}

\begin{document}

\title{Simulations of the Sterile Neutrino Oscillations with a Crossing-Width Term}

\author{Qiankang Wang$^a$}

\author{Dayun Qiu$^b$}

\author{Yi-Lei Tang$^a$}
\thanks{tangylei@mail.sysu.edu.cn}

\affiliation{$^a$School of Physics, Sun Yat-Sen University, Guangzhou 510275, China}
\affiliation{$^b$School of Physics and Astronomy, Sun Yat-sen University (Zhuhai Campus), Zhuhai 519082, China}

\begin{abstract}
In this paper, we present an algorithm to generate the collider events of the GeV-scale oscillating sterile neutrinos with the ready-made event generation tools in the case that the crossing-widths among the nearly-degenerate fermionic fields arise. We prove the validity of our algorithm, and adopt some tricks for practical calculations. The formulations of the particle oscillation processes are also improved in the framework of the quantum field theory, offering us the ability to simulate the flying distances of the oscillating intermediate sterile neutrinos while regarding them as the internal lines in the Feynmann diagrams.

\end{abstract}

\pacs{}

\keywords{}

\maketitle
\tableofcontents

\section{Introduction}

The type-I see-saw mechanism\cite{Minkowski:1977sc, Glashow:1979nm, Gell-Mann:1979vob, Mohapatra:1979ia, Yanagida:1980xy, Schechter:1980gr}, as perhaps the most prominent theory to endow the active neutrinos with masses, has lots of variations. The original version of the type-I see-saw mechanism accommodates the extremely heavy Majorana ``right-handed'', or ``sterile'' neutrinos well beyond the TeV scale to suppress the masses of the active neutrinos into sub-eV scale, if a moderate Yukawa coupling value $\sim0.01$-$1$ is assumed. Definitely no hope is left for us to verify such a heavy object at a terrestrial collider. However, in some variants of the type-I see-saw model\cite{Wyler:1982dd, Mohapatra:1986bd, Malinsky:2005bi},   
 if two nearly mass-degenerate sterile neutrinos co-exist to cancel most of their contributions to the light neutrino masses, opportunities of their productions at the TeV-scale colliders arise as their masses can be reduced to the GeV-scale\cite{Antusch:2015mia, Antusch:2016ejd, Antusch:2022ceb}. Such a mass degeneracy is usually guaranteed by the $U(1)_{\text{L}}$ symmetry, and the deviation of the mass degeneracy as well as the active neutrino masses are both proportional to the Majorana mass terms of the sterile neutrinos violating the $U(1)_{\text{L}}$ symmetry. In this case, such a pair of the nearly-degenerate sterile neutrinos can be regarded as one ``pseudo-Dirac'' neutrino.

The possibility to find out the (sub-)TeV scale sterile-neutrino has long been discussed in the literature (See \cite{Drewes:2013gca, Drewes:2015iva, Deppisch:2015qwa, Antusch:2016ejd, Antusch:2016vyf, Das:2017pvt, Antusch:2017pkq, Chun:2017spz, Chun:2017spz, Das:2018hph, Cvetic:2018elt, Bondarenko:2018ptm, Antusch:2018bgr, Drewes:2019fou, Cvetic:2019rms, Das:2019fee, Drewes:2024bla, Drewes:2024pad, Batra:2023mds, Batra:2023ssq, Dib:2016wge, Dib:2017vux, Dib:2017iva, Gao:2021one, Feng:2021eke, Gu:2022muc, Gu:2022nlj, Bai:2022lbv} for the theoretical discussions, and see Ref.~\cite{Abdullahi:2022jlv} for a review of the experimental results, and for the references therein). The key signals to distinguish the Majorana and Pseudo-Dirac neutrinos are the same-sign and different-sign lepton pairs discovered in the events, which are called the lepton number violating (LNV) and the lepton number conserving (LNC) events, respectively. If the mass difference between the pseudo-Dirac sterile neutrino pair becomes comparable with the width, then oscillations between the particle and the anti-particle arise so both the LNV and LNC signals appear\cite{Anamiati:2016uxp, Antusch:2017ebe, Antusch:2020pnn, Antusch:2022ceb, Antusch:2022ceb, Antusch:2022hhh, Antusch:2023nqd, Antusch:2023jsa, Antusch:2024otj}.

There are two algorithms to numerically simulate such an oscillation. Intuitively one can produce a nearly ``on-shell'' pseudo-Dirac sterile neutrino while oscillating and decaying it through some ``patches'' of a simulator\cite{Antusch:2022ceb}. If the lifetime of the sterile neutrino is long enough, the length of the ``displaced vertice'' can also be evaluated\cite{Cottin:2018kmq, Jana:2018rdf, Deppisch:2015qwa, Liu:2019ayx, Chiang:2019ajm, Urquia-Calderon:2023dkf}. Another way is to appoint the correct masses, widths and the rotational matrix parameters of the mass-eigenstate sterile neutrinos within the parameter configuration files, and let the unpatched simulator straightforwardly interferes the resonances of each Breit-Wigner resonances. All the tools here are ready-made, however the detailed information of the oscillation processes are shrouded.

In the literature usually the ``crossing-width'' term is also neglected. The crossing terms of the widths indicate the imaginary parts of the self-energy diagrams with two different particles as their external lines\cite{Pilaftsis:1997dr, Cacciapaglia:2009ic, Boyanovsky:2017esz, Qiu:2023zfr}. These might not be neglected if the two particles are nearly-degenerated. We have discussed such a term in our Ref.~\cite{Qiu:2023zfr} as a bosonic example. In the fermionic case, the nearly-degenerate pseudo-Dirac sterile neutrino pair is a good example\cite{Pilaftsis:1997jf, Pilaftsis:2003gt, Bray:2007ru}, although in the usual inverse or linear see-saw models, such a term automatically vanishes up to the lowest order due to the approximate lepton number symmetry. However, This term can in turn arise if one tries to extend the model. For an example, in Ref.~\cite{Falkowski:2011xh, Escudero:2016ksa, Falkowski:2017uya, Tang:2016sib, Bian:2018mkl}, a dark sector interacting with the sterile neutrino is introduced to break down the lepton number explicitly, generating the crossing-width term which can be resummed into the sterile neutrino propagators and might play important roles in finding the sterile neutrino signals at a collider.

In this paper, we aim at simulating the pseudo-Dirac sterile neutrino events with the crossing-width terms added. All the tools we rely on are ready-made without any patches setup or codes modified. We show that in order to sum up the crossing-width terms, a two-step diagonalizing algorithm should be adopted. During the simulation processes the sterile neutrinos are regarded as internal propagators rather than some ``on-shell'' intermediate objects, but the output event files leave us sufficient information for a stand-alone simulation of the oscillation processes and the distance of the displaced vertices according to our understanding of the quantum field theory. As an example, the ratio between the LNV and the LNC cross sections are calculated and presented for a paradigmatic $p p \rightarrow W^{\pm *} \rightarrow \mu^{\pm} N \rightarrow \mu \mu j j$ channel, and some crucial technical details, such as the algorithm to acquire the crossing-width terms without calculating the Feynmann diagrams by hand, the techniques to compute the sterile neutrino oscillations, will be clarified. 

\section{the Reference Model}

In this paper we consider a model accommodating both the sterile neutrino and the dark sector. The sterile neutrino can appear to communicate between the dark and the standard model (SM) sectors.  Besides, the pseudo-Dirac sterile neutrino field $N_D$ composed of two independent Weyl fields $N_L$ and $N_R$, that is to say
\begin{eqnarray}
    N_D = \left( \begin{array}{c}
         N_L  \\
         i \sigma^2 N_R^* 
    \end{array} \right), \label{ND_Def}
\end{eqnarray}
one Majorana fermion $\chi$ as well as one real-scalar field $\phi$ are also introduced as the elements of the dark sector, charged minus under the introduced dark $Z_2$ symmetry to keep the stability of the dark matter. The general Lagrangian, according to the Ref.~\cite{Bian:2018mkl}, is given by
\begin{eqnarray}
    \mathcal{L} &=& \frac{1}{2} \overline{\chi} (i \gamma^{\mu} \partial_{\mu} - m_{\chi} ) \chi + \overline{N_D} (i \gamma^{\mu} \partial_{\mu} - m_{N_D}) N_D + \frac{1}{2} (\partial^{\mu} \phi \partial_{\mu} \phi - m_{\phi}^2 \phi^2) \nonumber \\
    &+& (\mu_1 \overline{N_D^C} P_L N_D + \mu_2 \overline{N_D^C} P_R N_D + \text{h.c.} ) + \frac{\lambda}{4} \phi^4 + \lambda_{h \phi} \phi^2 H^{\dagger} H \nonumber \\
    &+& (y_{\chi D} \overline{\chi} N_D \phi + i y_{\chi D 5} \overline{\chi} \gamma^5 N_D \phi + y_{N i} \overline{N} P_L l_i \cdot H + y_{NC i} \overline{N^C} P_L l_i \cdot H \nonumber \\
    &+& \text{h.c.} ) + \mathcal{L_{\text{SM}}}, \label{L_Full}
\end{eqnarray}
where $y_{\chi D, \chi D 5}$, $\lambda_{\phi, \phi H}$ are coupling constants, and $m_{N_D}$, $m_{\phi, \chi}$, $\mu_{1,2}$ are the mass parameters. The charge conjugate operator $\psi^C$ means $i \gamma^2 \gamma^0 \bar{\psi}$ for any Dirac four-spinor $\psi$. As usual, the SM left-handed leptons and the Higgs doublet are given by
\begin{eqnarray}
    l_i = \left( \begin{array}{c} 
    \nu_i \\
    e_{L i}^- \end{array} \right), ~~ 
    H = \left( \begin{array}{c}
         G^+  \\
         \frac{v + h + i G^0}{\sqrt{2}}
    \end{array} \right),
\end{eqnarray}
where $\nu_i$ and $e_{L i}$ are the left-handed neutrino and charged lepton fields and $i=1,2,3$, corresponding to the $e$, $\mu$, $\tau$, respectively. $G^{0, \pm}$ are the Goldstone degrees of freedom, $h$ is the SM Higgs field, and the SM vacuum expectation value (VEV) $v = 246 \text{ GeV}$. 

In this paper, we focus on the sterile-neutrino sector, while neglect all the other detailed discussions like the dark matter phenomenology, the cosmological phase transitions as well as the corresponding gravitational wave signals, etc.. Parameters like $\lambda$, $\lambda_{h \phi}$ are thereby ignored, however they still appear in (\ref{L_Full}) for the completion of the general form of the Lagrangion.

\section{the Standard Step Diagonalization Processes}

After the $H$ acquires the VEV, the two Weyl components of the $N_D$ is no longer degenerate and will mix with the SM neutrinos. Take apart $N_D$ by (\ref{ND_Def}) in (\ref{L_Full}), we acquire the mass matrix
\begin{eqnarray}
    \mathcal{L} \supset - \frac{1}{2} \left( (i \sigma^2 \nu_i^*)^{\dagger}, ~  (i \sigma^2 N_L^*)^{\dagger}, ~ (i \sigma^2 N_R^*)^{\dagger} \right) \left( \begin{array}{ccc}
        0_{3 \times 3} & m_{D 3 \times 1} & m_{D 3 \times 1}^{\prime} \\
        m_{D 3 \times 1}^T & \mu_1 & m_R \\
        m_{D 3 \times 1}^{\prime T} & m_R & \mu_2
    \end{array} \right) \left( \begin{array}{c}
         \nu_i  \\
         N_L \\
         N_R
    \end{array} \right) + \text{h.c.}, \label{Complete_Seesaw}
\end{eqnarray}
where $(m_{D 3 \times 1})_i = y_{N i} v$, $(m_{D 3 \times 1}^{\prime})_i = y_{NC i} v$. If one would like a sub-TeV scale sterile neutrino which takes some opportunities to be detected in a terrestrial collider, an approximate $U(1)_L$ symmetry is required, therefore without loss of generality, $m_{D}^{\prime} \ll m_D$, $\mu_{1,2} \ll m_R$. When $m_{D}^{\prime} = 0$, (\ref{Complete_Seesaw}) indicates the ``inverse'' ($\mu_1=0$) see-saw or ``linear'' ($\mu_2 =0$) see-saw model. Therefore we adopt the $m_D^{\prime} = 0$, and assume that the sterile neutrino only mixes with the second generation of neutrino $\nu_{\mu}$ for the simplicity of our calculations. These simplify (\ref{Complete_Seesaw}) into
\begin{eqnarray}
    \mathcal{L} \supset - \frac{1}{2} \left( (i \sigma^2 \nu_{\mu}^*)^{\dagger}, ~  (i \sigma^2 N_L^*)^{\dagger}, ~ (i \sigma^2 N_R^*)^{\dagger} \right) \mathcal{M}_{N \nu} \left( \begin{array}{c}
         \nu_{\mu}  \\
         N_L \\
         N_R
    \end{array} \right) + \text{h.c.}, \label{Partial_Seesaw}
\end{eqnarray}
where
\begin{eqnarray}
    \mathcal{M}_{N\nu} = \left( \begin{array}{ccc}
        0 & m_{D \mu} & 0 \\
        m_{D \mu}^T & \mu_1 & m_{N_D} \\
        0 & m_{N_D} & \mu_2
    \end{array} \right). \label{Mass_Matrix_Original}
\end{eqnarray}

Generally (\ref{Mass_Matrix_Original}) is a complex symmetric matrix satisfying $\mathcal{M}_{N\nu}^T = \mathcal{M}_{N\nu}$, and one can diagonalize it by working on a Hermitian matrix
\begin{eqnarray}
    \mathcal{M}_{N \nu 2} = \mathcal{M}_{N \nu} \mathcal{M}_{N \nu}^{\dagger} = \mathcal{M}_{N \nu}^{T} \mathcal{M}_{N \nu}^{*}.
\end{eqnarray}
There should exist a unitary matrix $U$, so that
\begin{eqnarray}
    U^{T} \mathcal{M}_{N \nu 2}^T U^* = U^{\dagger} \mathcal{M}_{N \nu 2} U = \left( \begin{array}{ccc}
        \hat{m}_1^2 & 0 & 0 \\
        0 & \hat{m}_2^2 & 0 \\
        0 & 0 & \hat{m}_3^2
    \end{array} \right).
\end{eqnarray}
Define 
\begin{eqnarray}
    \hat{\mathcal{M}}_{\nu} = U^{\dagger}\mathcal{M}_{N\nu}U^{*},
\end{eqnarray}
so
\begin{eqnarray}
    U^{\dagger} \mathcal{M}_{N \nu 2} U = \hat{\mathcal{M}}_{\nu} \hat{\mathcal{M}}_{\nu}^{*} = \hat{\mathcal{M}}_{\nu}^* \hat{\mathcal{M}}_{\nu} = U^{T} \mathcal{M}_{N \nu 2}^T U^*.
\end{eqnarray}
Since $U^{\dagger} \mathcal{M}_{N \nu 2} U $ is diagonal, $v_1 = (1,0,0)^T$, $v_2 = (0,1,0)^T$ and $v_3 = (1,0,0)^T$ are its eigenvectors satisfying $\hat{\mathcal{M}}_{\nu} \hat{\mathcal{M}}_{\nu}^{*} v_i = \hat{\mathcal{M}}_{\nu}^{*} \hat{\mathcal{M}}_{\nu} v_i = U^{\dagger} \mathcal{M}_{N \nu 2} U v_i = \hat{m}_i^2 v_i$ (Here, the Einstein rules are temporally abandoned on index $i$). If $m_{1,2,3}^2$ are different nonzero numbers, and notice that $\hat{\mathcal{M}}_{\nu}^{*} \hat{\mathcal{M}}_{\nu} \hat{\mathcal{M}}_{\nu}^{*} v_i = \hat{\mathcal{M}}_{\nu}^{*} (\hat{\mathcal{M}}_{\nu} \hat{\mathcal{M}}_{\nu}^{*}) v_i = \hat{\mathcal{M}}_{\nu}^{*} U^{\dagger} \mathcal{M}_{N \nu 2} U v_i = \hat{m}_i^2 \hat{\mathcal{M}}_{\nu}^{*} v_i$, therefore $\hat{\mathcal{M}}_{\nu}^* v_i$ is also the eigenvector of the $\hat{\mathcal{M}}_{\nu}^{*} \hat{\mathcal{M}}_{\nu} = U^{\dagger} \mathcal{M}_{N \nu 2} U$ with the eigenvalue $m_i^2$, making $\hat{\mathcal{M}}_{\nu}^* v_i \propto v_i$ since $m_{1,2,3}^2$ are three different numbers. Therefore, $v_{1,2,3}$ are also the eigenvectors of the $\hat{\mathcal{M}}_{\nu}^{*}$, so $\hat{\mathcal{M}}_{\nu}^{*}$, thereby $\hat{\mathcal{M}}_{\nu}$ automatically becomes diagonalized.

It is easy to realize that the squared absolute values of the $\hat{\mathcal{M}}_{\nu}$'s diagonal elements are $m_{1,2,3}^2$ respectively, however the complex phase angles remain undetermined. Therefore
\begin{eqnarray}
    \hat{\mathcal{M}_{\nu}} = \left(
    \begin{array}{ccc}
       \hat{m_{1}} e^{i\delta_{1}} & 0&0 \\
        0 &\hat{m_{2}}e^{i\delta_{2}}&0\\
        0&0&\hat{m_{3}e^{i\delta_{3}}}
    \end{array}
    \right),
\end{eqnarray}
where $\delta_{1,2,3}$ are the three Majorana phases. To kill these phases, we utilize 
\begin{eqnarray}
    V_{\delta} = \left( \begin{array}{ccc}
        e^{i \frac{\delta_1}{2}} & 0 & 0 \\
        0 & e^{i \frac{\delta_2}{2}} & 0 \\
        0 & 0 & e^{i \frac{\delta_3}{2}} \\
    \end{array} \right)
\end{eqnarray}
for the final results
\begin{eqnarray}
    \mathcal{M}_{\nu} = V_{\delta}^{\dagger}\hat{\mathcal{M}}_{\nu}V_{\delta}^{*}=\left(
    \begin{array}{ccc}
       \hat{m}_{1}  &0&0  \\
        0 &\hat{m}_{2}&0\\
        0&0&\hat{m}_{3}
    \end{array}
    \right). \label{Eigen_Results}
\end{eqnarray}
Now, written in the Dirac four-component spinors, the Lagrangian becomes
\begin{eqnarray}
    \mathcal{L} \supset -\frac{1}{2} \sum_{i=1,2,3} \overline{\mathcal{N}_i} \mathcal{M}_{\nu ii} \mathcal{N}_i,
\end{eqnarray}
where $\mathcal{N}_i = \mathcal{N}_i^C$  is the majorana spinor field 
\begin{eqnarray}
    \mathcal{N}_i =\left(
    \begin{array}{c}
         \chi_i  \\
         i \sigma^2 \chi_i^{*}
    \end{array}\right), \label{Ni_Def}
\end{eqnarray}
where
\begin{eqnarray}
    \chi_i = \left[ (UV_{\delta})^{\dagger}
    \left(\begin{array}{c}
        \nu_{\mu}\\
        N_{L}\\
        N_{R}
    \end{array}\right) \right]_i .\label{diagrela1}
\end{eqnarray}

\section{the Second Step Diagonalization Processes, the Resummation of the Imaginary Parts from the Self-Energy Diagrams}

The first-step diagonalization processes work on the two-dimensional Weyl spinors, while for the FeynRules\cite{Alloul:2013bka} package, only the four-dimensional self-conjugate Dirac-form spinors (\ref{Ni_Def}) can be input. The definitions of the particles require the information of the widths, which originate from the imaginary parts of the self-energy diagrams.

\begin{figure}
    \centering
    \includegraphics[width=0.3\linewidth]{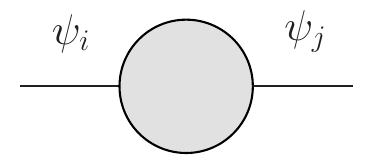}
    \caption{(Crossed) self-energy diagram connecting the $\psi_{i}$ and $\psi_{j}$ fermions.}
    \label{CrossedSelfEnergy}
\end{figure}

Let us begin with a general case of $n$ nearly-degenerate Majorana 4-spinor fields $\psi_{1}$, $\psi_{2}$, \dots, $\psi_{n}$ with the nearly unified approximate mass $m_{1,2,\dots n} \approx m$. Ref.~\cite{Gonchar:2006xv} had established the Breit-Wigner propagator of a single fermion, and now we have to derive into the multiple fermion cases. Denote the results of the diagram in Fig.~\ref{CrossedSelfEnergy} as $\Sigma_{i j}(\slashed{p}) + \Sigma_{i j}^5(\slashed{p}) \gamma^5$ from $\psi_{i}$ to $\psi_{j}$, where all the terms without a $\gamma^5$ are attributed into $\Gamma_{i j}$, and all the terms with a $\gamma^5$ are collected to form the $\Sigma_{i j}^5(\slashed{p}) \gamma^5$.

Extending the results in Ref.~\cite{Gonchar:2006xv} into the multiple fermion field cases with the $\gamma^5$ terms considered\cite{Bray:2007ru} by summing over the imaginary parts of the $\Sigma_{i j}(\slashed{p})$ and $\Sigma^5_{i j}(\slashed{p})$ in the $\Sigma_{i j}(\slashed{p}) + i \Sigma_{i j}^5(\slashed{p}) \gamma^5$, in which the real parts are nearly canceled by the counter terms. One then acquires
\begin{eqnarray}
    & & \left[\Delta_{F ij}^{\text{R}}(p)\right]_{n \times n} =\frac{i}{\slashed{p}I_{n\times n}-\mathcal{M}+i{\rm Im}[\Sigma(\slashed{p})] - {\rm Im}[\Sigma^5(\slashed{p})] \gamma^5} \nonumber \\
    &=&\Lambda^{+}\frac{i}{\sqrt{p^{2}}I_{n\times n}-\mathcal{M}+i{\rm Im}[\Sigma(\sqrt{p^2})] + {\rm Im}[\Sigma^5(\sqrt{p^{2}})] \gamma^5} \nonumber \\
    & & +\Lambda^{-}\frac{i}{-\sqrt{p^{2}}I_{n\times n}-\mathcal{M}+i{\rm Im}[\Sigma(-\sqrt{p^2})] + {\rm Im}[\Sigma^5(-\sqrt{p^{2}})] \gamma^5}, \label{Resummed_Propgator}
\end{eqnarray}
where $\Delta_{F ij}^{\text{R}}(p)$ are the resummed propagators connecting field $\psi_i$ with $\overline{\psi}_j$, the $\Lambda^{\pm}=\frac{1}{2}(1\pm \frac{\slashed{p}}{\sqrt{p^{2}}})$ are the two projector operators, and the $\Sigma^{(5)}(\slashed{p}) = \left[ \Sigma_{i j}^{(5)}(\slashed{p}) \right]_{n \times n}$ are the collections of the $\Sigma^{(5)}_{i j}(\slashed{p})$'s. $\mathcal{M}$ denotes ${\rm diag}[m_1, m_2, \dots, m_n]$. When the slashed four-vector parameters in the $\Sigma_{i j}^{(5)}(\slashed{p})$ are replaced with the scalar $\pm \sqrt{p^2}$, this means that all the $\slashed{p}$ terms in the expansions of the $\Sigma_{i j}(\slashed{p})$ are replaced with the $\pm \sqrt{p^{2}}$. Since we concentrate near the ``shell'' $p^2 \approx m^2$, the parameters $\sqrt{p^2}$ in all ${\rm Im}(\Sigma^{(5)}_{i j}(\sqrt{p^2})$'s can be replaced with the $m$ as a good approximation. Denote
\begin{equation}
    \Gamma_{ij} = 2{\rm Im}[ \Sigma_{ij}(m) ],~~\Gamma_{ij}^5 = 2{\rm Im}[ \Sigma_{ij}^5(m) ].
\end{equation}
In the (\ref{Resummed_Propgator}), it is easy to realize that the first term dominates near the shell $\sqrt{p^2} \approx m$, and it can be confirmed that
\begin{eqnarray}
    & & \frac{1}{\sqrt{p^{2}}I_{n\times n}-\mathcal{M}+i\frac{\Gamma}{2} + \frac{\Gamma^5}{2} \gamma^5} \nonumber \\
    &=& (A+BA^{-1}B)^{-1} + (B+AB^{-1}A)^{-1} \gamma^5, \label{CompletePropagatorsWith5}
\end{eqnarray}
where $\Gamma^{(5)} = \left[ \frac{\Sigma_{i j}^{(5)}(m)}{2} \right]_{n \times n}$ are the matrix forms, and we also set
\begin{eqnarray}
    A &=& \sqrt{p^{2}}I_{n\times n}-\mathcal{M}+i\frac{\Gamma}{2}, \nonumber \\
    B &=& -\frac{\Gamma^5}{2}.
\end{eqnarray}

When both $\det(A+BA^{-1}B)=0$ and $\det(B+AB^{-1}A)=0$ in the case of a particular $\sqrt{p^2}$, the solution of the linear equations $(A+BA^{-1}B) x = 0$ or $(B+AB^{-1}A) x = 0$ for the unknown vector $x$ indicates the ``mass-eigenstate'' in combination of the $\psi_{i}$'s, since a pole is encountered at this $\sqrt{p^2}$ and $x$ in (\ref{CompletePropagatorsWith5}). To find out all the poles, notice that
\begin{eqnarray}
    & &\det(A+BA^{-1}B) = \det(A B^-1 A + B)A^{-1} B = \det[(A+i B) B^{-1} (A-i B) A^{-1} B] \nonumber \\
    & & = \det(A^2+B^2 + iBA-iAB) \det A^{-1} , \\
    & & \det(B+AB^{-1}A) = \det(A^2+B^2 + iBA-iAB) \det B^{-1},
\end{eqnarray}
so finding the poles in (\ref{CompletePropagatorsWith5}) can be attributed to solving the equation
\begin{eqnarray}
    \det(A+iB)\det(A-iB) = \det (A^2+B^2 + iBA-iAB) = 0. \label{Expected_Function}
\end{eqnarray}
If all the subdominant $\Gamma_{ij}^{(5)}$'s are in the similar scale, and we would like to only keep their linear contributions to the corrections of the solution positions, then $B^2$ can be safely neglected. For the $i(BA-AB)$ terms, notice that  $BA-AB \approx \frac{-1}{2} (\mathcal{M} \Gamma^5 - \Gamma^5 \mathcal{M})$, and the nearly degenerate $\mathcal{M}$ can be denoted by $\mathcal{M} = \bar{m} I_{n \times n} + \text{diag}[\delta m_1, \delta m_2, \dots, \delta m_n] = \bar{m} I_{n \times n} \delta \mathcal{M}$, where $\bar{m} = \frac{\text{tr}\mathcal{M}}{n}$, and thereby $\delta m_i = m_i-\bar{m}$, $\delta \mathcal{M} = \mathcal{M}-\bar{m} I_{n \times n}$. Then $BA-AB \approx \frac{-1}{2} (\delta \mathcal{M} \Gamma^5 - \Gamma^5 \delta \mathcal{M})$ and can also be neglected since the $\delta \mathcal{M}$ is subdominant due to the nearly degeneracy of the $\mathcal{M}$. Therefore, only the $A^2$ is preserved in the (\ref{Expected_Function}), and all the $\Sigma^5_{i j}$ terms are neglected. In such a case, it would be better for us to neglect the $\Sigma^5$ from the beginning. Therefore (\ref{Resummed_Propgator}) becomes
\begin{eqnarray}
    \left[\Delta_{F ij}^{\text{R}}(p)\right]_{n \times n} &\approx& \Lambda^{+}\frac{i}{\sqrt{p^{2}}I_{n\times n}-\mathcal{M}+i \frac{\Gamma}{2} }+\Lambda^{-}\frac{i}{-\sqrt{p^{2}}I_{n\times n}-\mathcal{M}+i \frac{\Gamma}{2} } \nonumber \\
    &=& \frac{i}{\slashed{p}I_{n\times n}-(\mathcal{M}-i\Gamma/2)}.
\end{eqnarray}

Unlike the Ref.~\cite{Pilaftsis:1997jf, Pilaftsis:2003gt, Bray:2007ru} which reverse the propagator matrices straightforwardly, we alternatively utilize a matrix $V$ to diagonalize the propagators $\left[\Delta_{F ij}^{\text{R}}(p)\right]_{n \times n}$ in the form
\begin{eqnarray}
    V \left[ \Delta_{F ij}^{\text{R}}(p) \right]_{n \times n} V^{-1} &=& V \frac{i}{\slashed{p}I_{n\times n}-V(\mathcal{M}-i\Gamma/2)V^{-1}} V^{-1} = \frac{i}{\slashed{p}I_{n\times n}-V(\mathcal{M}-i\Gamma/2)V^{-1}} \nonumber \\
    &=& {\rm diag}[\frac{i(\slashed{p}+m_{1}^{\prime})}{p^{2}-m_{1}^{\prime 2}+im_{1}^{\prime} \Gamma_{1}},~ \frac{i(\slashed{p}+m_{2}^{\prime})}{p^{2}-m_{2}^{\prime 2}+im_{2}^{\prime} \Gamma_{2}},~ \dots], \label{Dressed_Propagator_Diagonalized}
\end{eqnarray}
which is equivalent to diagonalizing the non-Hermitian ``mass-matrix'' $\mathcal{M}^{\prime} = \mathcal{M} - i \frac{\Gamma}{2}$ into
\begin{eqnarray}
    V \mathcal{M}^{\prime} V^{-1} = V(\mathcal{M}-i\frac{\Gamma}{2}) V^{-1} ={\rm diag}[m_{1}^{\prime}-\frac{i\Gamma_{1}^{\prime}}{2},~m_{2}^{\prime}-\frac{i\Gamma_{2}^{\prime}}{2},~\dots]. \label{diag_3}
\end{eqnarray}
This tells us that it is critical for us to compute all the ``crossed-widths'' $\Gamma_{i j}$ when $i \neq j$ besides all the familiar diagonal elements $\Gamma_{i i}$. As a summary, we have shown that after the traditional diagonlization processes upon the Weyl spinors like (\ref{diagrela1})
, one has to further pick up all the nearly-degenerate fields before transforming them into the 4-spinor forms, and then compute all the (crossed-)widths to formulate the non-Hermitian ``mass-matrix'' $\mathcal{M} - i \frac{\Gamma}{2}$ for the third-stage diagonalizion process as we have presented in (\ref{diag_3}).

Now let us draw back to our $\nu$-$N_{L,R}$ system. If we sort the eigenvalues in (\ref{Eigen_Results}) to an ascending order, that is to say, $\hat{m}_1 < \hat{m}_2 < \hat{m}_3$, generally $\hat{m}_1$ should be tiny since $m_{D \mu} \ll m_R$, indicating the ``active'' light neutrino, and $\hat{m}_2 \approx \hat{m}_3 \approx m_{N_D}$ indicating the two nearly-degenerate degrees of freedom. Therefore at this stage we only have to consider the $\Gamma_{22, 33, 23}$ to diagonalize the non-Hermitian ``mass-matrix'' $\mathcal{M}_{\nu}^{\prime}$,
\begin{eqnarray}
    \mathcal{M}_{\nu}' = \mathcal{M}_{\nu}-i\frac{\Gamma_{ij}}{2}= \left(
        \begin{array}{ccc}
            \hat{m}_{1}&0&0\\
            0&\hat{m}_{2}-i\frac{\Gamma_{22}}{2}&-i\frac{\Gamma_{23}}{2}\\
            0&-i\frac{\Gamma_{23}}{2}&\hat{m}_{3}-i\frac{\Gamma_{33}}{2}
        \end{array}
    \right). \label{Mass_Gamma}
\end{eqnarray}
with a complex orthogonal matrix $V$ into
\begin{eqnarray}
    \mathcal{M}_{f} = V^{T}\mathcal{M}_{\nu}'V = \left(
        \begin{array}{ccc}
            m_{\mathcal{N}_1^{\prime}} = \hat{m}_1 &0&0\\
            0&m_{\mathcal{N}_2^{\prime}}-\frac{i}{2}\Gamma_{\mathcal{N}_2^{\prime}}&0\\
            0&0&m_{\mathcal{N}_3^{\prime}}-\frac{i}{2}\Gamma_{\mathcal{N}_3^{\prime}}
        \end{array}
    \right). \label{Diag_2stage_Sterile}
\end{eqnarray}
Here, the active light neutrino mass remains intact so $m_{\mathcal{N}_1^{\prime}} = \hat{m}_1$, and only the $\mathcal{N}_{2,3}$ fields are affected.

Rigorously speaking, such a diagonalization process is only an operation upon the resummed propagators (\ref{Resummed_Propgator}), however in order for a convenient numerical simulation with the popular event generators like the MadGraph\cite{Alwall:2014hca, Frederix:2018nkq}, effectively (\ref{Diag_2stage_Sterile}) can be understood by rotating the $\mathcal{N}_i$ defined in (\ref{Ni_Def}) into
\begin{eqnarray}
    \mathcal{N}_i^{\prime} = V_{i j}^T \mathcal{N}_j, ~~\overline{\mathcal{N}_i^{\prime}} = V_{i j}^T \overline{\mathcal{N}}_j. \label{NiPrime_Def}
\end{eqnarray}
Here $\overline{\mathcal{N}_i^{\prime}} \neq \mathcal{N}_i^{\prime \dagger} \gamma^0$, unlike the usual definition of a 4-spinor hatted with a ``bar''. All the corresponding mass terms and coupling terms are also rotated at the same time. It also seems weird that  the $\mathcal{N}_i^{\prime}$ are no longer the ``self-charge-conjugate'' 4-spinors since
\begin{eqnarray}
    \mathcal{N}_i^{\prime} = \left( \begin{array}{c}
         V_{i j}^T \chi_j  \\
         V_{i j}^T \chi_j^C 
    \end{array} \right). 
\end{eqnarray}
However, when input into a numerical package, $\mathcal{N}_i^{\prime}$'s should all be regarded as the Majorana 4-spinors in order for a self-consistent simulation considering the diagonalized propagator (\ref{Dressed_Propagator_Diagonalized}). A similar trick in the bosonic case has been applied in our previous work Ref.~\cite{Qiu:2023zfr}.



Before the close of this section, we have to note that if we switch off both the $y_{\chi_{D}}$ and $y_{\chi_{D5}}$ to acquire a pure see-saw model, the $\Gamma_{23}$ in (\ref{Mass_Gamma}) vanishes and $\Gamma_{22} = \Gamma_{33}$ due to the symmetry exchanging the $\mathcal{N}_2$ and $\mathcal{N}_3$ in the diagonalized Lagrangian, which is a $Z_2$ subgroup of the approximated $U(1)_L$ symmetry. These had been confirmed by our numerical calculations. Therefore one does not have to concern about the cross widths connecting $\mathcal{N}_{2,3}$ in the usual inverse or linear see-saw model, although such a term might exist and should be carefully considered in such a reference model described by (\ref{L_Full}).

\section{Sterile neutrino oscillations, principles and simulations} \label{Oscillation_Section}

It is convenient to regard the $\mathcal{N}_{2,3}^{\prime}$ always as the intermediate propagators during the simulation processes, however if $\Gamma_{\mathcal{N}_2^{\prime}},\Gamma_{\mathcal{N}_3^{\prime}}$ are pretty small, displaced vertices might arise as well as the inevitable oscillation effects. In the literature, the simulations of the oscillations are accomplished by separately calculating the emission of the on-shell sterile neutrinos and their decay/oscillation processes within a hybrid framework of both the quantum field theory and the quantum mechanics. In this paper, we attempt to simulate the oscillations regarding $\mathcal{N}_{2,3}^{\prime}$ as internal propagators in the complete framework of the quantum field theory. Therefore, we have to combine the two seemingly contradicting frameworks, and try to understand the nearly-on-shell particle propagation and oscillation processes in a unified diagramatic framework.

In the literature, if one does not want to formulate the oscillation issues in the form of the ``quantum mechanics''\cite{Boyanovsky:2014una, Drewes:2016gmt}, usually the oscillation theory within the quantum field theory framework involves regarding both the incoming and the outgoing particles as wave packets\cite{Naumov:2010um, Akhmedov:2017xxm, Antusch:2023nqd}. This is not so convenient since all collider simulators are based upon the calculations of the S-matrices with the assumption that the wave functions and the interaction locations of the particles pervade all through the infinite space-time volume, therefore all possible four-dimensional interval between the generation and decay of the sterile neutrinos are summed over. Then, our task is to separate out the portion within the S-matrix of a particular space-time interval in which the intermediate sterile neutrino subsists. Usually, when some nearly-degenerate objects propagate as the intermediate particles, the usual $S$-matrix can be formulated into
\begin{eqnarray}
    S = \dots \int \frac{d^4 p}{(2 \pi)^4} (2 \pi)^4 \delta^{(4)}(p_1 + p_2 + \dots - p) \sum_i A_i \Delta_{(F) i}(p) (2 \pi)^4 \delta^{(4)} ( p - p_1^{\prime} - p_2^{\prime} - \dots) \dots, \label{S_Matrix}
\end{eqnarray}
where the two $\delta$ functions indicate the energy-momentum conservation laws with the $p_1$, $p_2$, $\dots$, $p_1^{\prime}$, $p_2^{\prime}$, $\dots$ the four-dimensional momentums of the particles connecting the $\Delta_{(F) i}$ propagators. The different couplings related with the particle $i$ are all attributed into the parameter $A_i$. $\Delta_{(F)i}(p)$'s are the propagators, being $\Delta_{i}(p) = \frac{i}{p^2 - m_i^2 + i m_i \Gamma_i}$ for bosons and $\Delta_{F i}(p) = \frac{i(\slashed{p}+m)}{p^2 - m_i^2 + i m_i \Gamma_i}$ for fermions. Of course, (\ref{S_Matrix}) shrouds the oscillation information, unless we notice that the two $\delta$-functions there originate from
\begin{eqnarray}
    \int d^4 x_1 e^{i (p_1 + p_2 + \dots - p) \cdot x_1} &\rightarrow& (2 \pi)^4 \delta^ {(4)}(p_1 + p_2 + \dots - p), \nonumber \\
    \int d^4 x_2 e^{i (p - p_1^{\prime} - p_2^{\prime} - \dots) \cdot x_2 } &\rightarrow& (2 \pi)^4 \delta^ {(4)}(p_1 + p_2 + \dots - p),
\end{eqnarray}
where $x_1$ and $x_2$ can be regarded as the two vertices that the oscillating particles generate and decay, so $x_1 - x_2 = \Delta x$ becomes the space-time interval that the intermediate particles subsist. Therefore,  (\ref{S_Matrix}) can be formulated into
\begin{eqnarray}
    & & S = \dots \int \frac{d^4 p}{(2 \pi)^4} d^4 x_1 d^4 x_2 e^{i (p_1 + p_2 + \dots - p) \cdot x_1} \sum_i A_i \Delta_{(F) i}(p) e^{i (p - p_1^{\prime} - p_2^{\prime} - \dots) \cdot x_2 } \dots \nonumber \\
    &=& \dots \int \frac{d^4 p}{(2 \pi)^4} d^4 \Delta x \frac{d^4 \overline{x}}{2}  e^{i (p_1 + p_2 + \dots) \cdot (\overline{x} + \frac{\Delta x}{2}) - i(-p_1^{\prime} - p_2^{\prime} - \dots) \cdot (\overline{x} - \frac{\Delta x}{2})} e^{ -i p \cdot \Delta x} \sum_i A_i \Delta_{(F) i}(p) \dots  \nonumber \label{S_Matrix_Interval} \\
    &=& \dots \int \frac{d^4 p}{(2 \pi)^4} d^4 \Delta x \frac{(2 \pi)^4}{2} \delta^4 (p_1 + p_2 + \dots - p_1^{\prime} - p_2^{\prime} - \dots)   e^{i (\frac{p_1 + p_2 + \dots + p_1^{\prime} + p_2^{\prime} + \dots}{2} - p) \cdot (\Delta x)}  \sum_i A_i \Delta_{(F) i}(p) \dots, \nonumber \\
    &=&\dots \int \frac{d^4 p}{(2 \pi)^4} d^4 \Delta x \frac{(2 \pi)^4}{2} \delta^4 (p_1 + p_2 + \dots - p_1^{\prime} - p_2^{\prime} - \dots)   e^{i (p_1 + p_2 + \dots - p) \cdot (\Delta x)}  \sum_i A_i \Delta_{(F) i}(p) \dots. \label{S_Matrix_Interval} 
\end{eqnarray}
where $\overline{x} = \frac{x_1+x_2}{2}$, which can be integrated out to leave us only the space-time interval $\Delta x$ information. 

Since in this paper, we are discussing the oscillations of the fermions, so $\Delta_{F i}(p) = \frac{i(\slashed{p}+m)}{p^2 - m_i^2 + i m_i \Gamma_i}$, therefore
\begin{eqnarray}
    & & S = \dots \int \frac{d^4 p}{(2 \pi)^4} d^4 \Delta x \frac{(2 \pi)^4}{2} \delta^4 (p_1 + p_2 + \dots - p_1^{\prime} - p_2^{\prime} - \dots) e^{i (p_1 + p_2 + \dots) \cdot (\Delta x)} \nonumber \\
    &\times& \sum_i e^{-i p \cdot \Delta x}   A_i \frac{i (p \cdot \gamma + m_i)}{p^2 - m_i^2 + i m_i \Gamma_i} \dots. \label{S_xbar_Int}
\end{eqnarray}

Integrate out the $\vec{p}$ at first is convenient. Notice that
\begin{eqnarray}
    & & \int \frac{d^4 p}{(2 \pi)^4} \frac{ie^{-i p \cdot \Delta x}}{p^2 - m_i^2 + i m_i \Gamma_i} = \int \frac{dp^0}{2 \pi} e^{-i p^0 \Delta x^0} \int \frac{2 \pi \vec{p}^2 \sin \theta d\vec{p} d \theta}{(2 \pi)^3} \frac{i e^{|\vec{p}| |\Delta \vec{x}| \cos \theta}}{(p^0)^2 - \vec{p}^2 - m_i^2 + i m_i \Gamma_i} \nonumber \\
    &=& \int \frac{dp^0}{2 \pi} e^{-i p^0 \Delta x^0} \int \frac{2 \pi \vec{p}^2 \sin \theta d|\vec{p}| d \theta}{(2 \pi)^3} \frac{i e^{|\vec{p}| |\Delta \vec{x}| \cos \theta}}{(p^0)^2 - \vec{p}^2 - m_i^2 + i m_i \Gamma_i} \nonumber \\
    &=& \int \frac{dp^0}{2 \pi} e^{-i p^0 \Delta x^0} \int_0^{\infty} \frac{2 \pi \vec{p}^2 d|\vec{p}|}{(2 \pi)^3 \left( i |\vec{p}| |\Delta \vec{x}| \right) } \frac{i \left( e^{i |\vec{p}| |\Delta \vec{x}|} - e^{-i |\vec{p}| |\Delta \vec{x}|} \right)}{(p^0)^2 - \vec{p}^2 - m_i^2 + i m_i \Gamma_i} \nonumber \\
    &=& \int \frac{dp^0}{2 \pi} e^{-i p^0 \Delta x^0} \int_{-\infty}^{\infty} \frac{q dq}{(2 \pi)^2 |\Delta \vec{x}|} \frac{e^{i q |\Delta \vec{x}|} }{(p^0)^2 - q^2 - m_i^2 + i m_i \Gamma_i} \nonumber \\
    &=& \int \frac{dp^0}{2 \pi} e^{-i p^0 \Delta x^0} \frac{2 \pi i}{(2 \pi)^2 |\Delta \vec{x}|} \frac{e^{i \sqrt{(p^0)^2- m_i^2 + i m_i \Gamma_i} |\Delta \vec{x}| }}{2},
\end{eqnarray}
where in the last step the residue of $q = \sqrt{(p^0)^2- m_i^2 + i m_i \Gamma_i}$ is picked up. With this result (\ref{S_xbar_Int}) becomes
\begin{eqnarray}
    & & S = \dots \int dp^0 d^4 \Delta x \frac{(2 \pi)^4}{2} \delta^4 (p_1 + p_2 + \dots - p_1^{\prime} - p_2^{\prime} - \dots) e^{i (p_1 + p_2 + \dots) \cdot (\Delta x)} \nonumber \\
    &\times& \sum_i \frac{i}{8 \pi^2}  A_i \left( p^0 \gamma^0 - \frac{\vec{\gamma}}{i} \cdot \vec{\nabla}_{\Delta \vec{x}}  + m_i \right) \frac{e^{i \sqrt{(p^0)^2 - m_i^2 + i m_i \Gamma_i} |\Delta \vec{x}| - i p^0 \Delta x^0}}{2 |\Delta \vec{x}|} \dots, \label{S_Matrix_pNoInted}
\end{eqnarray}
where $\vec{\nabla}_{\Delta \vec{x}}$ means the gradient operator about the parameter $\Delta \vec{x}$. 

Notice that usually it is the space interval which can be reconstructed by the detectors, since tracing the particles' flitting processes is beyond the time resolution of the usual practical equipments, we therefore further integrate out the time-related parameter $\Delta x^0$ and $p^0$ in (\ref{S_Matrix_pNoInted}),
\begin{eqnarray}
    & & S = \dots \int d^3 \Delta \vec{x} \frac{(2 \pi)^4}{2} \delta^4 (p_1 + p_2 + \dots - p_1^{\prime} - p_2^{\prime} - \dots) e^{i (\vec{p}_1 + \vec{p}_2 + \dots) \cdot\vec{\Delta x}} \nonumber \\
    &\times& \left. \sum_i \frac{i}{4 \pi }  A_i \left( p^0 \gamma^0 - \frac{\vec{\gamma}}{i}  \cdot \vec{\nabla}_{\Delta \vec{x}} + m_i \right) \frac{e^{i \sqrt{(p^0)^2 - m_i^2 + i m_i \Gamma_i} |\Delta \vec{x}| }}{2 |\Delta \vec{x}|} \right|_{p^0 = p_1^0 + p_2^0 + \dots} \dots, \nonumber \\
    & & \approx \dots \int d^3 \Delta \vec{x} \frac{(2 \pi)^4}{2} \delta^4 (p_1 + p_2 + \dots - p_1^{\prime} - p_2^{\prime} - \dots) e^{i (\vec{p}_1 + \vec{p}_2 + \dots) \cdot\vec{\Delta x}} \nonumber \\
    &\times& \left. \sum_i \frac{i}{4 \pi |\Delta \vec{x}|}  A_i \left( p^0 \gamma^0 - \sqrt{(p^0)^2 - m_i^2 + i m_i \Gamma_i} \frac{\Delta \vec{x}}{|\Delta \vec{x}|} \cdot \vec{\gamma} + m_i \right) \frac{e^{i \sqrt{(p^0)^2 - m_i^2 + i m_i \Gamma_i} |\Delta \vec{x}| }}{2} \right|_{p^0 = p_1^0 + p_2^0 + \dots} \dots. \label{S_Matrix_pInted}
\end{eqnarray}
Here we applied
\begin{eqnarray}
    & & \vec{\nabla}_{\Delta \vec{x}} \frac{e^{i \sqrt{(p^0)^2 - m_i^2 + i m_i \Gamma_i} |\Delta \vec{x}| } }{2 |\Delta \vec{x}|} = \frac{e^{i \sqrt{(p^0)^2 - m_i^2 + i m_i \Gamma_i} |\Delta \vec{x}|} \left(  i \sqrt{(p^0)^2 - m_i^2 + i m_i \Gamma_i} \Delta\vec{x} - \frac{\Delta\vec{x}}{|\Delta\vec{x}|} \right) }{2 |\Delta \vec{x}|^2} \nonumber \\
    & \approx & \frac{e^{i \sqrt{(p^0)^2 - m_i^2 + i m_i \Gamma_i} |\Delta \vec{x}|}  i \sqrt{(p^0)^2 - m_i^2 + i m_i \Gamma_i} \Delta\vec{x} }{2 |\Delta \vec{x}|^2},
\end{eqnarray}
where the last step is only valid for the macroscopic $\Delta x$ so that $\sqrt{(p^0)^2 - m_i^2 + i m_i \Gamma_i} \Delta\vec{x} \gg 1$.

Since we mainly concern about the distance $| \Delta \vec{x}|$ between the vertices, and pay less attention on its direction, it is then convenient to integrate out the angular parameters of the $\Delta \vec{x}$.
Notice that for a three-dimensional vector $\vec{q}$ and any function $f(|\vec{q}|)$
\begin{eqnarray}
    & & \int  e^{i \Delta \vec{x} \cdot \vec{q}} f(|\vec{q}|) d^3 \Delta \vec{x} = \int 2 \pi |\Delta \vec{x}|^2 \sin \theta d \theta d|\Delta x| e^{i |\Delta \vec{x}| |\vec{q}| \cos \theta} f(|\vec{q}|) \nonumber \\
    &=& 2 \pi \int_0^{\infty} |\Delta \vec{x}|^2 \frac{-1}{i |\Delta \vec{x}| |\vec{q}|} \left( e^{-i |\Delta \vec{x}| |\vec{q}|} - e^{i |\Delta \vec{x}| |\vec{q}|} \right)  f(|\vec{q}|) d|\Delta \vec{x}|,
\end{eqnarray}
and
\begin{eqnarray}
    & & \int d^3 \Delta \vec{x} e^{i \Delta \vec{x} \cdot \vec{q}} \Delta\vec{x} f(|\vec{q}|) = \int 2 \pi |\Delta \vec{x}|^2 \sin \theta d \theta d|\Delta x| e^{i |\Delta \vec{x}| |\vec{q}| \cos \theta} \left( |\Delta \vec{x}| \cos \theta \frac{\vec{q}}{|\vec{q}|} \right) f(|\vec{q}|) \nonumber \\
    =& & 2 \pi \int_0^{\infty} |\Delta \vec{x}|^2 \frac{-1}{i |\Delta \vec{x}| |\vec{q}|} \frac{\vec{q}}{\vec{q}^2} \left( e^{-i |\Delta \vec{x}| |\vec{q}|} (1 + i |\Delta \vec{x}| |\vec{q}|) - e^{i |\Delta \vec{x}| |\vec{q}| } (1 - i |\Delta \vec{x}| |\vec{q}|) \right) d|\Delta \vec{x}| f(|\vec{q}|) \nonumber \\
    \approx & & 2 \pi \int_0^{\infty} |\Delta \vec{x}|^2 \frac{-1}{i} \frac{\vec{q}}{\vec{q}^2} i \left( e^{-i |\Delta \vec{x}| |\vec{q}|} + e^{i |\Delta \vec{x}| |\vec{q}| } \right) d|\Delta \vec{x}| f(|\vec{q}|),
\end{eqnarray}
where again, we applied the macroscopic $\Delta \vec{x}$ approximation $|\vec{q}| |\Delta \vec{x}| \gg 1$. Also notice that we are discussing a bunch of nearly on-shell oscillating particles, so $\sqrt{(p^0)^2 - m_i^2} \approx |\vec{p}_1 + \vec{p}_2 + \dots|$ should be satisfied. In this case, $e^{i (\sqrt{(p^0)^2 - m_i^2} + |\vec{p}_1 + \vec{p}_2 + \dots|) |\Delta \vec{x}|}$ rapidly oscillates even when $|\Delta \vec{x}|$ changes a little, so its macroscopic contribution can be neglected, and therefore
\begin{eqnarray}
    & & S \approx \dots \int d |\Delta \vec{x}| \frac{(2 \pi)^4}{2} \delta^4 (p_1 + p_2 + \dots - p_1^{\prime} - p_2^{\prime} - \dots) e^{-i |\vec{p}_1 + \vec{p}_2 + \dots| |\Delta \vec{x}|} \frac{-2 \pi}{i |\vec{p}_1 + \vec{p}_2 + \dots|} \nonumber \\
    &\times& \left. \sum_i \frac{i}{4 \pi}  A_i \left( p^0 \gamma^0 - \sqrt{(p^0)^2 - m_i^2 + i m_i \Gamma_i} \frac{\vec{p}_1 + \vec{p}_2 + \dots}{|\vec{p}_1 + \vec{p}_2 + \dots|} \cdot \vec{\gamma} + m_i \right) \frac{e^{i \sqrt{(p^0)^2 - m_i^2 + i m_i \Gamma_i} |\Delta \vec{x}| }}{2} \right|_{p^0 = p_1^0 + p_2^0 + \dots} \dots. \nonumber \\
    & \approx & \dots \int d |\Delta \vec{x}| \frac{(2 \pi)^4}{2} \delta^4 (p_1 + p_2 + \dots - p_1^{\prime} - p_2^{\prime} - \dots) e^{-i |\vec{p}_1 + \vec{p}_2 + \dots| |\Delta \vec{x}|} \frac{-2 \pi}{i |\vec{p}_1 + \vec{p}_2 + \dots|} \nonumber \\
    &\times& \left. \sum_i \frac{i}{4 \pi}  A_i \left( p^0 \gamma^0 - (\vec{p}_1 + \vec{p}_2 + \dots) \cdot \vec{\gamma} + m_i \right) \frac{e^{i \sqrt{(p^0)^2 - m_i^2 + i m_i \Gamma_i} |\Delta \vec{x}| }}{2} \right|_{p^0 = p_1^0 + p_2^0 + \dots} \dots, \label{S_Final}
\end{eqnarray}
where during the last step, we applied the ``nearly on-shell'' approximations $\sqrt{(p^0)^2 - m_i^2 + i m_i \Gamma_i} \approx \sqrt{(p^0)^2 - m_i^2} \approx |\vec{p}_1 + \vec{p}_2 + \dots|$ in the narrow width conditions.

The integrand in (\ref{S_Final}) can be interpreted that the probability of the space interval to become $|\Delta \vec{x}|$ is proportional to
\begin{eqnarray}
    P_{|\Delta \vec{x}|} \propto \left| \cdots \sum_i A_i i(p^{\prime} \cdot \gamma + m_i) e^{i \sqrt{(p^0)^2 - m_i^2 + i m_i \Gamma_i} |\Delta \vec{x}| } \cdots \right|^2, \label{Oscilation_Basic}
\end{eqnarray}
where $p^{\prime} = p_1 + p_2 + \dots$ is the approximate four-momentum of the propagating nearly-degenerate particles. The oscillation effects originate from $e^{i \sqrt{(p^0)^2 - m_i^2 + i m_i \Gamma_i} |\Delta \vec{x}| }$, in which the tiny differences between the $m_i$'s induce the macroscopic $\sin \left( \frac{m_i^2 - m_j^2}{p_0} |\Delta \vec{x}| \right)$  or  $\cos \left( \frac{m_i^2 - m_j^2}{p_0} |\Delta \vec{x}| \right)$ oscillation terms.

Now we are ready to simulate the distance of the displaced vertices of the $\mathcal{N}_{1, 2}^{\prime}$. Based upon (\ref{Diag_2stage_Sterile}) 
one can write down the ``propagators'' of the $\mathcal{N}_{1,2}^{\prime}$
\begin{eqnarray}
    ``\langle 0 | T[N_i \overline{N_j}] | 0 \rangle'' \rightarrow \delta_{ij} \frac{i (\slashed{p} + m_i)}{p^2 - m_i^2 + i m_i \Gamma_i}. \label{Basic_Propagators}
\end{eqnarray}
Compare (\ref{Basic_Propagators}) with terms appeared in (\ref{Oscilation_Basic}), one can clearly realize that the denominator $p^2 - m_i^2 + i m_i \Gamma_i$ should be replaced with the $e^{i \sqrt{(p^0)^2 - m_i^2 + i m_i \Gamma_i} |\Delta \vec{x}| }$ factor. Then the general steps to generate the displaced vertex distance $|\Delta \vec{x}|$ can be followed as
\begin{itemize}
    \item Take (\ref{Basic_Propagators}) into (\ref{S_Matrix}) for a normal calculation of the $S$-matrix for the total probability when the intermediate $\mathcal{N}_{1,2}^{\prime}$ are nearly on-shell
    \item Replace all the $\frac{i}{p^2 - m_i^2 + i m_i \Gamma_i}$ with the corresponding $e^{i \sqrt{(p^0)^2 - m_i^2 + i m_i \Gamma_i}}$ factors to formulate the probability function $P_{|\Delta \vec{x}|}$.
    \item Generate the $|\Delta \vec{x}|$ randomly with the probability distribution function $P_{|\Delta \vec{x}|}$ for the displaced vertex distance value.
\end{itemize}

Manually enumerating all the possible diagrams involving different incoming and outgoing asymptotic states are extremely cumbersome, and it is also inconvenient to hack into the codes of the numerical tools to modify the propagator forms. In this paper, we point out that sometimes, if regarded as the oscillation between the ``particle'' $N_D$ and the ``anti-particle'' $\overline{N}_D$, the properties of both the sterile neutrinos' creation and decay vertices can be inferred by the identities of the by-product particles. For an example, if the sterile neutrino is decayed from a $W^{-}$ and therefore a $\mu^{-}$ is detected, and finally a $\mu^{+}$ is decayed from the sterile neutrino, it will then be inferred that both the objects at the creation and decay points are the anti-sterile neutrinos, and therefore it is the $\langle 0 | T[N_D \overline{N_D}] | 0 \rangle $ that contributes into the S-matrices to generate such an event. On the other hand, when the sterile neutrino is decayed from a $W^{-}$ boson so that a $\mu^{-}$ arises at this point, while the sterile neutrino finally decays into a $\mu^{-}$, we in turn infer that an anti-sterile neutrino is generated before it oscillates into a sterile neutrino and decay, making use of the $\langle 0 | T[N_D^C \overline{N_D}] | 0 \rangle $ propagator, where $N_D^c = i \gamma^2 \gamma^0 \overline{N_D}^{\dagger}$.

Define
\begin{eqnarray}
    V_L &=& U V_{\delta} V, \\
    V_R &=& (U V_{\delta})^* V,
\end{eqnarray}
one can calculate $\langle 0 | T[N_D \overline{N_D}] | 0 \rangle $, $\langle 0 | T[N_D \overline{N_D^C}] | 0 \rangle $, $\langle 0 | T[N_D^C \overline{N_D}] | 0 \rangle $, and $\langle 0 | T[N_D^C \overline{N_D^C}] | 0 \rangle $ according to (\ref{diagrela1}),(\ref{NiPrime_Def}). Omitting the tiny values in $V_{L,R}$ elements involving the ``active'' neutrinos, we have
\begin{gather}
    N_{D}=P_{L}[(V_{L})_{22} \mathcal{N}_{2}^{\prime} +(V_{L})_{23}\mathcal{N}_{3}^{\prime}]+P_{R}[(V_{R})_{32} \mathcal{N}_{2}^{\prime}+(V_{R})_{33} \mathcal{N}_{3}^{\prime}], \\
    \overline{N_{D}}=[(V_{L})_{32}\overline{\mathcal{N}_{2}^{\prime}}+(V_{L})_{33} \overline{\mathcal{N}_{3}^{\prime}}]P_{L} + [(V_{R})_{22} \overline{\mathcal{N}_{2}^{\prime}} + (V_{R})_{23} \overline{\mathcal{N}_{3}^{\prime}}]P_{R},\\
    N_{D}^{c} =P_{R}[(V_{R})_{22} \mathcal{N}_{2}^{\prime}+(V_{R})_{23} \mathcal{N}_{3}^{\prime}] + P_{L}[(V_{L})_{32} \mathcal{N}_{2}^{\prime} + (V_{L})_{33} \mathcal{N}_{3}^{\prime}], \\
    \overline{N_{D}^{c}}=[(V_{L})_{22}\overline{\mathcal{N}_{2}^{\prime}} + (V_{L})_{23} \overline{\mathcal{N}}_{3}^{\prime}] P_{L} + [(V_{R})_{32} \overline{\mathcal{N}_{2}^{\prime}} + (V_{R})_{33} \overline{\mathcal{N}_{3}^{\prime}}] P_{R}.
\end{gather}
In this section we show all the propagators as
\begin{eqnarray}
    \begin{aligned}
    & \langle 0|T[N_{D}\overline{N_{D}}]|0\rangle \\
    &\rightarrow [(V_{L})_{22}(V_{L})_{32}\frac{im_{\mathcal{N}_2^{\prime}}}{p^{2}-m_{\mathcal{N}_2^{\prime}}^{2}+im_{\mathcal{N}_2^{\prime}}\Gamma_{\mathcal{N}_2^{\prime}}}+(V_{L})_{23}(V_{L})_{33}\frac{im_{\mathcal{N}_3^{\prime}}}{p^{2}-m_{\mathcal{N}_3^{\prime}}^{2}+im_{\mathcal{N}_3^{\prime}}\Gamma_{\mathcal{N}_3^{\prime}}}]P_{L}\\
    &+[(V_{L})_{22}(V_{R})_{22}\frac{i\slashed{p}}{p^{2}-m_{\mathcal{N}_2^{\prime}}^{2}+im_{\mathcal{N}_2^{\prime}}\Gamma_{\mathcal{N}_2^{\prime}}}+(V_{L})_{23}(V_{R})_{23}\frac{i\slashed{p}}{p^{2}-m_{\mathcal{N}_3^{\prime}}^{2}+im_{\mathcal{N}_3^{\prime}}\Gamma_{\mathcal{N}_3^{\prime}}}]P_{R}\\
    &+[(V_{L})_{32}(V_{R})_{32}\frac{i\slashed{p}}{p^{2}-m_{\mathcal{N}_2^{\prime}}^{2}+im_{\mathcal{N}_2^{\prime}}\Gamma_{\mathcal{N}_2^{\prime}}}+(V_{L})_{33}(V_{R})_{33}\frac{i\slashed{p}}{p^{2}-m_{\mathcal{N}_3^{\prime}}^{2}+im_{\mathcal{N}_3^{\prime}}\Gamma_{\mathcal{N}_3^{\prime}}}]P_{L}\\
    &+[(V_{R})_{22}(V_{R})_{32}\frac{im_{\mathcal{N}_2^{\prime}}}{p^{2}-m_{\mathcal{N}_2^{\prime}}^{2}+im_{\mathcal{N}_2^{\prime}}\Gamma_{\mathcal{N}_2^{\prime}}}+(V_{R})_{23}(V_{R})_{33}\frac{im_{\mathcal{N}_3^{\prime}}}{p^{2}-m_{\mathcal{N}_3^{\prime}}^{2}+im_{\mathcal{N}_3^{\prime}}\Gamma_{\mathcal{N}_3^{\prime}}}]P_{R} = G_1(p), \\
    \end{aligned} \label{NDND}
\end{eqnarray}
\begin{equation}
    \begin{aligned}
        & \langle 0|T[N_{D}\overline{N_{D}^{c}}]|0\rangle\\
         &\rightarrow [(V_{L})_{22}(V_{R})_{32}\frac{i\slashed{p}}{p^{2}-m_{\mathcal{N}_2^{\prime}}^{2}+im_{\mathcal{N}_2^{\prime}}\Gamma_{\mathcal{N}_2^{\prime}}}+(V_{L})_{23}(V_{R})_{33}\frac{i\slashed{p}}{p^{2}-m_{\mathcal{N}_3^{\prime}}^{2}+im_{\mathcal{N}_3^{\prime}}\Gamma_{\mathcal{N}_3^{\prime}}}]P_{R}\\
         &\ \ \ +[(V_{L})_{22}^{2}\frac{im_{\mathcal{N}_2^{\prime}}}{p^{2}-m_{\mathcal{N}_2^{\prime}}^{2}+im_{\mathcal{N}_2^{\prime}}\Gamma_{\mathcal{N}_2^{\prime}}}+(V_{L})_{23}^{2}\frac{im_{\mathcal{N}_3^{\prime}}}{p^{2}-m_{\mathcal{N}_3^{\prime}}^{2}+im_{\mathcal{N}_3^{\prime}}\Gamma_{\mathcal{N}_3^{\prime}}}]P_{L}\\
         &\ \ \ +[(V_{R})_{32}^{2}\frac{im_{\mathcal{N}_2^{\prime}}}{p^{2}-m_{\mathcal{N}_2^{\prime}}^{2}+im_{\mathcal{N}_2^{\prime}}\Gamma_{\mathcal{N}_2^{\prime}}}+(V_{R})_{33}^{2}\frac{im_{\mathcal{N}_3^{\prime}}}{p^{2}-m_{\mathcal{N}_3^{\prime}}^{2}+im_{\mathcal{N}_3^{\prime}}\Gamma_{\mathcal{N}_3^{\prime}}}]P_{R}\\
         &\ \ \ +[(V_{R})_{32}(V_{L})_{22}\frac{i\slashed{p}}{p^{2}-m_{\mathcal{N}_2^{\prime}}^{2}+im_{\mathcal{N}_2^{\prime}}\Gamma_{\mathcal{N}_2^{\prime}}}+(V_{R})_{33}(V_{L})_{23}\frac{i\slashed{p}}{p^{2}-m_{\mathcal{N}_3^{\prime}}^{2}+im_{\mathcal{N}_3^{\prime}}\Gamma_{\mathcal{N}_3^{\prime}}}]P_{L}=G_{2}(p),
    \end{aligned} \label{NDNDc}
\end{equation}
\begin{equation}
    \begin{aligned}
       &\langle 0|T[N_{D}^{c}\overline{N_{D}^{c}}]|0\rangle\\
        &\rightarrow [(V_{R})_{32}(V_{R})_{22}\frac{im_{\mathcal{N}_2^{\prime}}}{p^{2}-m_{\mathcal{N}_2^{\prime}}^{2}+im_{\mathcal{N}_2^{\prime}}\Gamma_{\mathcal{N}_2^{\prime}}}+(V_{R})_{33}(V_{R})_{23}\frac{im_{\mathcal{N}_3^{\prime}}}{p^{2}-m_{\mathcal{N}_3^{\prime}}^{2}+im_{\mathcal{N}_3^{\prime}}\Gamma_{\mathcal{N}_3^{\prime}}}]P_{R}\\
        &\ \ \ +[(V_{R})_{22}(V_{L})_{22}\frac{i\slashed{p}}{p^{2}-m_{\mathcal{N}_2^{\prime}}^{2}+im_{\mathcal{N}_2^{\prime}}\Gamma_{\mathcal{N}_2^{\prime}}}+(V_{R})_{23}(V_{L})_{23}\frac{i\slashed{p}}{p^{2}-m_{\mathcal{N}_3^{\prime}}^{2}+im_{\mathcal{N}_3^{\prime}}\Gamma_{\mathcal{N}_3^{\prime}}}]P_{L}\\
        &\ \ \ +[(V_{L})_{32}(V_{R})_{32}\frac{i\slashed{p}}{p^{2}-m_{\mathcal{N}_2^{\prime}}^{2}+im_{\mathcal{N}_2^{\prime}}\Gamma_{\mathcal{N}_2^{\prime}}}+(V_{L})_{33}(V_{R})_{33}\frac{i\slashed{p}}{p^{2}-m_{\mathcal{N}_3^{\prime}}^{2}+im_{\mathcal{N}_3^{\prime}}\Gamma_{\mathcal{N}_3^{\prime}}}]P_{R}\\
        &\ \ \ +[(V_{L})_{32}(V_{L})_{22}\frac{im_{\mathcal{N}_2^{\prime}}}{p^{2}-m_{\mathcal{N}_2^{\prime}}^{2}+im_{\mathcal{N}_2^{\prime}}\Gamma_{\mathcal{N}_2^{\prime}}}+(V_{L})_{33}(V_{L})_{23}\frac{im_{\mathcal{N}_3^{\prime}}}{p^{2}-m_{\mathcal{N}_3^{\prime}}^{2}+im_{\mathcal{N}_3^{\prime}}\Gamma_{\mathcal{N}_3^{\prime}}}]P_{L}=G_{3}(p)\\
    \end{aligned} \label{NDcNDc}
\end{equation}
\begin{equation}
    \begin{aligned}
       &\langle0|T[N_{D}^{c}\overline{N_{D}}]|0\rangle\\
        &\rightarrow [(V_{R})_{22}(V_{L})_{32}\frac{i\slashed{p}}{p^{2}-m_{\mathcal{N}_2^{\prime}}^{2}+im_{\mathcal{N}_2^{\prime}}\Gamma_{\mathcal{N}_2^{\prime}}}+(V_{R})_{23}(V_{L})_{33}\frac{i\slashed{p}}{p^{2}-m_{\mathcal{N}_3^{\prime}}^{2}+im_{\mathcal{N}_3^{\prime}}\Gamma_{\mathcal{N}_3^{\prime}}}]P_{L}\\
        &\ \ \ +[(V_{R})_{22}^{2}\frac{im_{\mathcal{N}_2^{\prime}}}{p^{2}-m_{\mathcal{N}_2^{\prime}}^{2}+im_{\mathcal{N}_2^{\prime}}\Gamma_{1}}+(V_{R})_{23}^{2}\frac{im_{\mathcal{N}_3^{\prime}}}{p^{2}-m_{\mathcal{N}_3^{\prime}}^{2}+im_{\mathcal{N}_3^{\prime}}\Gamma_{\mathcal{N}_3^{\prime}}}]P_{R}\\
        &\ \ \ +[(V_{L})_{32}^{2}\frac{im_{\mathcal{N}_2^{\prime}}}{p^{2}-m_{\mathcal{N}_2^{\prime}}^{2}+im_{\mathcal{N}_2^{\prime}}\Gamma_{\mathcal{N}_2^{\prime}}}+(V_{L})_{33}^{2}\frac{im_{\mathcal{N}_3^{\prime}}}{p^{2}-m_{\mathcal{N}_3^{\prime}}^{2}+im_{\mathcal{N}_3^{\prime}}\Gamma_{\mathcal{N}_3^{\prime}}}]P_{L}\\
        &\ \ \ +[(V_{R})_{22}(V_{L})_{32}\frac{i\slashed{p}}{p^{2}-m_{\mathcal{N}_2^{\prime}}^{2}+im_{\mathcal{N}_2^{\prime}}\Gamma_{\mathcal{N}_2^{\prime}}}+(V_{L})_{33}(V_{R})_{23}\frac{i\slashed{p}}{p^{2}-m_{\mathcal{N}_3^{\prime}}^{2}+im_{\mathcal{N}_3^{\prime}}\Gamma_{\mathcal{N}_3^{\prime}}}]P_{R}=G_{4}(p),
    \end{aligned} \label{NDcND}
\end{equation}
and then one can replace all the $\frac{i}{p^2 - m_{\mathcal{N}_{2,3}^{\prime}}^2 + i m_{\mathcal{N}_{2,3}^{\prime}} \Gamma_{\mathcal{N}_{2,3}^{\prime}}}$ with $e^{i \sqrt{(p^0)^2 - m_{\mathcal{N}_i^{\prime}}^2 + i m_{\mathcal{N}_i^{\prime}} \Gamma_{\mathcal{N}_i^{\prime}}} |\Delta \vec{x}|}$ which can be estimated to be $e^{i \left[ \sqrt{(p^0)^2 - m_{\mathcal{N}_i^{\prime}}^2 } - \frac{ i m_{\mathcal{N}_i^{\prime}} \Gamma_{Ni} }{2 \sqrt{(p^0)^2 - m_{\mathcal{N}_i^{\prime}}^2}} \right] |\Delta \vec{x}|}$. Taking this into (\ref{Oscilation_Basic}) to compute the oscillation probability at the space interval $|\Delta \vec{x}|$ still involves complicated matrix calculations. To avoid this, notice that
\begin{eqnarray}
    I_{4 \times 4} = \frac{1}{2 m} \sum_{\lambda} [u(p, \lambda) \overline{u}(p, \lambda) - v(p, \lambda) \overline{v}(p, \lambda)], \label{Unit_4x4}
\end{eqnarray}
where $\lambda$ is the spin notation. Inserting (\ref{Unit_4x4}) at the two ``\dots'' symbols in (\ref{Oscilation_Basic}) does not change the results, and expanding $I_{4 \times 4} G_{t}(p) I_{4 \times 4} \propto \sum\limits_{\lambda} [u(p, \lambda) \overline{u}(p ,\lambda) - v(p, \lambda) \overline{v}(p, \lambda)] G_{t}(p) \sum\limits_{\rho} [u(p, \rho) \overline{u}(p, \rho) - v(p, \lambda) \overline{v}(p, \rho)] $, $t=1,2,3,4$ involves calculating the transition amplitudes within combinations of the sterile neutrinos and the anti-sterile neutrinos,
\begin{eqnarray}
    & & \overline{u}(p, \lambda) G_t(p) u(p, \rho),~\overline{u}(p, \lambda) G_t(p) v(p, \rho), \nonumber \\
    & & \overline{v}(p, \lambda) G_t(p) u(p, \rho),~\overline{v}(p, \lambda) G_t(p) v(p, \rho),
\end{eqnarray}
where $\lambda$, $\rho$ denote the spins. Since we are discussing the oscillation of the nearly-on-shell nearly-degenerate particles, therefore $\sqrt{p^2} \approx m_{\mathcal{N}_2^{\prime}} \approx m_{\mathcal{N}_3^{\prime}} \approx m_N=\frac{m_{\mathcal{N}_2^{\prime}}+m_{\mathcal{N}_2^{\prime}}}{2}$, so
\begin{eqnarray}
    & & \slashed{p} u(p, \lambda) \approx m_N u(p, \lambda),~\slashed{p} v(p, \lambda) \approx -m_N v(p, \lambda), \label{F1} \\
    & & \overline{u}(p, \lambda) \gamma^5 u(p, \rho) = \overline{v}(p, \rho) \gamma^5 v(p, \lambda) = \overline{v}(p, \lambda) \gamma^5 u(p, \rho) \nonumber \\
    & & = \overline{u}(p, \lambda) \gamma^5 v(p, \rho) = \overline{v}(p, \rho) u(p, \lambda) = \overline{u}(p, \lambda) v(p, \rho) = 0, \label{F2} \\
    & & \overline{u}(p, \lambda) u(p, \rho) = 2m\delta_{\lambda \rho},~\overline{v}(p,\rho)v(p,\lambda)=-2m\delta_{\rho \lambda}. \label{F3}
\end{eqnarray}
Then the oscillation terms can be easily discriminated. For an example, if one wants the probability that a sterile neutrino oscillates into a sterile neutrino while travelling a distance $|\Delta \vec{x}|$, he needs to calculate $\overline{u}(p, \lambda) G_1(p) u(p, \rho)$, or $\overline{v}(p, \rho) G_3(p) v(p, \lambda)$ before replacing all the $\frac{i}{p^2 - m_{\mathcal{N}_{2,3}^{\prime}}^2 + i m_{\mathcal{N}_{2,3}^{\prime}} \Gamma_{\mathcal{N}_{2,3}^{\prime}}}$ with $e^{i \left[ \sqrt{(p^0)^2 - m_{\mathcal{N}_i^{\prime}}^2 } - \frac{ i m_{\mathcal{N}_i^{\prime}} \Gamma_{\mathcal{N}_i^{\prime}} }{2 \sqrt{(p^0)^2 - m_{\mathcal{N}_i^{\prime}}^2}} \right] |\Delta \vec{x}|}$. Here, compared with the $\overline{u}(p, \lambda) G_1(p) u(p, \rho)$ annihilating a sterile neutrino state $u(p, \rho)$ while creating another sterile neutrino state $u(p, \lambda)$, $\overline{v}(p, \rho) G_3(p) v(p, \lambda)$ can be understood as annihilating an ``anti-particle'' $\overline{v}(p, \rho)$ of a charge conjugate field $N_D^c$,  while creating another ``anti-particle'' $v(p, \lambda)$, equivalent to a propagation from $u(p, \rho)$ to $u(p, \lambda)$ from the $N_D$ aspect. Moreover, since the momentum direction is reversed, one should also reverse the momentum sign to compute $\overline{v}(p, \rho) G_3(-p) v(p, \lambda)$. therefore both $\overline{u}(p, \lambda) G_1(p) u(p, \rho)$ and $\overline{v}(p, \rho) G_3(-p) v(p, \lambda)$ give
\begin{eqnarray}
    P_{N_D, \rho \rightarrow N_D, \lambda} \propto \delta_{\lambda \rho} [c_{1}c_{1}^{*}e^{-\frac{m\Gamma_{\mathcal{N}_2^{\prime}}}{E}t}+c_{2}c_{2}^{*}e^{-\frac{m\Gamma_{\mathcal{N}_3^{\prime}}}{E}t}+ 2 \text{Re}(c_{1}c_{2}^{*}e^{i(E_{2}-E_{1})t} )e^{-\frac{m(\Gamma_{\mathcal{N}_2^{\prime}}+\Gamma_{\mathcal{N}_3^{\prime}})}{2E}t}], \label{P_NDND}
\end{eqnarray}
where
\begin{gather}
    c_{1} =[(V_{L})_{22}+(V_{R})_{32}][(V_{L})_{32}+(V_{R})_{22}],\\
    c_{2} = [(V_{L})_{23}+(V_{R})_{33}][(V_{L})_{33}+(V_{R})_{23}].
\end{gather}
Besides, $t = \frac{|\Delta \vec{x}| p_0 }{\sqrt{(p^0)^2 - m_N^2}}$ can be understood as the effective ``flying time'' of the oscillating object, and $E_{i} = \sqrt{(p^0)^2 - m_N^2 + m_{\mathcal{N}_i^{\prime} }^2 }$ can be regarded as the ``energy'' of the $\mathcal{N}_{2,3}^{\prime}$ respectively, if their momentums are adjusted to be exactly the same.


Similarly, $P_{N_D, \rho \rightarrow \overline{N}_D, \lambda}$, $P_{\overline{N}_D , \rho \rightarrow N_D, \lambda}$, and $P_{\overline{N}_D, \rho \rightarrow \overline{N}_D, \lambda}$ can be acquired from $\overline{u}(p, \lambda) G_4(p) u(p, \rho) - \overline{v}(p, \rho) G_4(-p) v(p, \lambda)$, $\overline{u}(p, \lambda) G_2(p) u(p, \rho) - \overline{v}(p, \rho) G_2(-p) v(p, \lambda)$, $\overline{u}(p, \lambda) G_3(p) u(p, \rho)$ respectively. Here, $\overline{N}_D$ indicates the anti-sterile neutrino from the aspect of the $N_D$ field, and the ``minus'' sign within $\overline{u}(p, \lambda) G_4(p) u(p, \rho) - \overline{v}(p, \rho) G_4(-p) v(p, \lambda)$, $\overline{u}(p, \lambda) G_2(p) u(p, \rho) - \overline{v}(p, \rho) G_2(-p) v(p, \lambda)$ originate from the anti-commutation relation of the fermionic fields dduring the contraction operations.
The results are calculated to be
\begin{eqnarray}
        P_{N_D, \lambda_1 \rightarrow \overline{N}_D, \lambda_2}(t) &\propto& \delta_{\lambda \rho} [c_{3}c_{3}^{*}e^{-\frac{m\Gamma_{\mathcal{N}_2^{\prime}}}{E}t}+c_{4}c_{4}^{*}e^{-\frac{m\Gamma_{\mathcal{N}_3^{\prime}}}{E}t}+ 2 \text{Re} (c_{3}c_{4}^{*}e^{i(E_{2}-E_{1})t})e^{-\frac{m(\Gamma_{\mathcal{N}_2^{\prime}}+\Gamma_{\mathcal{N}_3^{\prime}})}{2E}t}], \label{P_NDNDBar} \\
        P_{\overline{N}_D, \lambda_1 \rightarrow  \overline{N}_D \lambda_2}(t) &\propto& \delta_{\lambda \rho} [c_{5}c_{5}^{*}e^{-\frac{m\Gamma_{\mathcal{N}_2^{\prime}}}{E}t}+c_{6}c_{6}^{*}e^{-\frac{m\Gamma_{\mathcal{N}_3^{\prime}}}{E}t}+2 \text{Re} (c_{5}c_{6}^{*}e^{i(E_{2}-E_{1})t})e^{-\frac{m(\Gamma_{\mathcal{N}_2^{\prime}}+\Gamma_{\mathcal{N}_3^{\prime}})}{2E}t}],  \label{P_NDBarNDBar} \\
        P_{\overline{N}_D, \lambda_1 \rightarrow N_D, \lambda_2}(t) &\propto&  \delta_{\lambda \rho} [c_{7}c_{7}^{*}e^{-\frac{m\Gamma_{\mathcal{N}_2^{\prime}}}{E}t}+c_{8}c_{8}^{*}e^{-\frac{m\Gamma_{\mathcal{N}_3^{\prime}}}{E}t}+ 2 \text{Re} (c_{7}c_{8}^{*}e^{i(E_{2}-E_{1})t})e^{-\frac{m(\Gamma_{\mathcal{N}_2^{\prime}}+\Gamma_{\mathcal{N}_3^{\prime}})}{2E}t}], \label{P_NDBarND}
\end{eqnarray}
where
\begin{eqnarray}
    c_{3} &=& [(V_{R})_{32}+(V_{L})_{22}]^{2}, \nonumber \\
    c_{4} &=& [(V_{R})_{33}+(V_{L})_{23}]^{2}, \nonumber \\
    c_{5} &=& [(V_{L})_{22}+(V_{R})_{32}][(V_{R})_{22}+(V_{L})_{32}], \nonumber \\
    c_{6} &=& [(V_{L})_{23}+(V_{R})_{33}][(V_{R})_{23}+(V_{L})_{33}], \nonumber \\
    c_{7} &=& [(V_{L})_{32}+(V_{R})_{22}]^{2}, \nonumber \\
    c_{8} &=& [(V_{L})_{33}+(V_{R})_{23}]^{2}.
\end{eqnarray}

(\ref{P_NDND}), (\ref{P_NDNDBar}), (\ref{P_NDBarNDBar}) and (\ref{P_NDBarND}) are the probabilities that a sterile neutrino oscillates into a sterile neutrino, a sterile neutrino oscillates into an anti-sterile neutrino, an anti-sterile neutrino oscillates into an anti-sterile neutrino, and an anti-sterile neutrino oscillates into a sterile neutrino, respectively. Practical evaluation requires a normalization of all these four functions before hand, and finally distribute the total probability of a particular event calculated by the usual S-matrix algorithm regarding the $\mathcal{N}_{2,3}^{\prime}$ as intermediate propagators into different $t = \frac{|\Delta \vec{x}| p_0 }{\sqrt{(p^0)^2 - m_N^2}}$ according to the probability distributions proportional to (\ref{P_NDND}), (\ref{P_NDNDBar}), (\ref{P_NDBarNDBar}) and (\ref{P_NDBarND}).

\section{Several Important Numerical Tricks}

In this paper, we aim at simulating the creation and decay processes of a pair of nearly-degenerate sterile neutrinos with the usual Monte-Carlo tools, like the MadGraph. However we found some subtleties during our attempts, and it is worthy for us to illustrate the details and the solutions. In this section, we are going to explain them.

\subsection{Cross-widths Calculations} \label{CrossWidthAlgorithm}

In (\ref{Mass_Gamma}), the capability to acquire the values of the diagonal $\Gamma_{22}$ and $\Gamma_{33}$ is the standard function implemented by all sorts of collider simulators, however the ``crossed'' $\Gamma_{23}$ draws no attention in the high energy physics community. In MadGraph, the widths of the designated particles can be computed before the formal beginning of the Monte-Carlo simulation processes, and no entrance for the cross-widths are left for the users.

In this paper, we propose an ingenious algorithm by introducing two auxillary ``dummy'' particles, one Majorana fermion $N^{\prime}$, and a real scalar $\phi^{\prime}$. The expedient Yukawa couplings
\begin{eqnarray}
    \mathcal{L}_{\text{dummy}} = y^{\prime}_i \phi^{\prime} \overline{N^{\prime}} \mathcal{N}_i + \text{h.c.}
\end{eqnarray}
are introduced.

Although event generators cannot compute the crossed $\mathcal{N}_2 \leftrightarrow \text{everything} \leftrightarrow \mathcal{N}_3$ processes, all of them are designed to simulate the collision processes with two initial particles. If the $\mathcal{N}_{2,3}$ appear as the nearly on-shell s-channel intermediators, effectively the crossed $\Gamma_{23}$ are hidden within the interference terms between the $\phi^{\prime} N^{\prime} \rightarrow \mathcal{N}_{2,3} \rightarrow \text{everything}$ diagrams.

Assign a unified temporary width to both $\mathcal{N}_{2,3}$, and appoint the center of value energy exactly at $\bar{m_{\mathcal{N}_{23}}} = \frac{\hat{m}_2 + \hat{m}_3}{2} \approx \hat{m}_2 \approx \hat{m}_3$, then compute the cross sections
\begin{eqnarray}
    & & \sigma_{N^{\prime} \phi^{\prime} \rightarrow \mathcal{N}_2 \rightarrow \text{everything}}, \nonumber \\
    & & \sigma_{N^{\prime} \phi^{\prime} \rightarrow \mathcal{N}_3 \rightarrow \text{everything}}, \nonumber \\
    & & \sigma_{N^{\prime} \phi^{\prime} \rightarrow (\mathcal{N}_{2} + \mathcal{N}_3) \rightarrow \text{everything}},
\end{eqnarray}

For each of the s-channel intermediators $\mathcal{N}_{2,3}$, a $\frac{\slashed{p} + m_{\mathcal{N}_{2,3}^{\prime}}}{p^2-m_{\mathcal{N}_{2,3}^{\prime}}^2}$ propagator arises, with the numerator $\slashed{p} + m_{N 2,3} = \sum\limits_{\lambda} u(p, \lambda) \overline{u}(p, \lambda)$ equivalent to the spin-sum of the $\mathcal{N}_{2,3}$, so
\begin{eqnarray}
    \sigma_{N^{\prime} \phi^{\prime} \rightarrow \mathcal{N}_2 \rightarrow \text{everything}} & \propto & \Gamma_{22},\nonumber \\
    \sigma_{N^{\prime} \phi^{\prime} \rightarrow \mathcal{N}_3 \rightarrow \text{everything}}& \propto & \Gamma_{33}, \nonumber \\
    \sigma_{N^{\prime} \phi^{\prime} \rightarrow (\mathcal{N}_{2} + \mathcal{N}_3) \rightarrow \text{everything}} & \propto & \Gamma_{22} + \Gamma_{33} + 2 \Gamma_{23},
\end{eqnarray}
Remember $\Gamma_{22}$ and $\Gamma_{33}$ can be easily computed by any event generator, so $\Gamma_{23}$ can be acquired from
\begin{eqnarray}
    \Gamma_{23} = \frac{  \sigma_{N^{\prime} \phi^{\prime} \rightarrow (\mathcal{N}_{2} + \mathcal{N}_3) \rightarrow \text{everything}} - \sigma_{N^{\prime} \phi^{\prime} \rightarrow \mathcal{N}_2 \rightarrow \text{everything}} - \sigma_{N^{\prime} \phi^{\prime} \rightarrow \mathcal{N}_3 \rightarrow \text{everything}} }{\sigma_{N^{\prime} \phi^{\prime} \rightarrow \mathcal{N}_2 \rightarrow \text{everything}}} \Gamma_{22}.
\end{eqnarray}

\subsection{the Cross Section Rescaling Algorithms} \label{Rescaling}

If an isolated particle $A$ with the mass $m_A$ is nearly on-shell as an s-channel intermediator, its width information $\Gamma_A$ is crucial for the event generation processes. The residue theorem tells us that the cross section can be estimated by
\begin{eqnarray}
    \sigma_{X \rightarrow A+Y \rightarrow Z+Y} &=& \int d \Pi_X d \Pi_Y d \Pi_Z \dots \frac{1}{\left( s - m^2 + i m \Gamma_A \right) \left( s - m^2 - i m \Gamma_A \right)^* } \dots \nonumber \\
    &=& \int (d\dots) \int_{s_1}^{s_2} ds \dots \frac{1}{\left( s - m^2 + i m \Gamma_A \right) \left( s - m^2 - i m \Gamma_A \right)^* } \dots \nonumber \\
    & \simeq & \int (d\dots) \int_{-\infty}^{\infty} ds \dots \frac{1}{\left( s - m^2 + i m \Gamma_A \right) \left( s - m^2 - i m \Gamma_A \right)^* } \dots \nonumber \\
    &=& \int (d\dots) \dots \frac{\pi}{m \Gamma_A} \dots = \sigma_{X \rightarrow A+Y} \frac{\Gamma_{A \rightarrow Z}}{\Gamma_A}, \label{StandardScale}
\end{eqnarray}
where $X$, $Y$ and $Z$ denote the particle sets other than the intermediator $A$ with the $d \Pi_X d \Pi_Y d \Pi_Z$ as their phase space measures. There must be some way to reorganize the phase space parameters to some other new parameters with the invariant mass $s$ of the momentum of the $A$ as a member. At the second line of the (\ref{StandardScale}), (d\dots) means the measures of the new phase space parameters other than $s$. If the range $[s_1, s_2]$ of the parameter $s$ to be integrated accommodates $m^2 \in [s_1, s_2]$ and the $\Gamma_A$ is small enough so that the residue at the $s = m^2 \pm i m \Gamma_A$ dominates the integration results, it is then convenient for us to pick up the residues of one of the poles to estimate the total cross section $\sigma_{X \rightarrow A+Y}$ regarding $A$ as an on-shell particle, multiplied by the branching ratio $\frac{\Gamma_{A \rightarrow Z}}{\Gamma_A}$.

Practically, if $\Gamma_A \ll 0.001 m_A$, numerically computing the $\sigma_{X \rightarrow A+Y \rightarrow Z+Y}$ straightforwardly confronts the problem of the numerical instabilities near the $s = m^2 \pm i m \Gamma_A$ poles. Therefore, appointing another larger $\Gamma_A^{\prime} \sim 0.005m_A \text{-} 0.05 m_A$ relaxes this problem with the price that the total cross section is rescaled by a factor, so that $\sigma_{X \rightarrow A+Y \rightarrow Z+Y} = \frac{\Gamma_A}{\Gamma^{\prime}_A} \sigma^{\prime}_{X \rightarrow A+Y \rightarrow Z+Y}$ where $\sigma^{\prime}_{X \rightarrow A+Y \rightarrow Z+Y}$ is the rescaled cross section calculated when the width of the $A$ is assigned to be $\Gamma_A^{\prime}$. If one does not care about the precise shape of the extremely narrow resonance near the $s = m_A^2$, the $\Gamma^{\prime}$ can be simply applied and the $\sigma^{\prime}_{X \rightarrow A+Y \rightarrow Z+Y}$ can be acquired through the rescaling algorithm.

Such an algorithm can be generalized when two nearly-degenerate $A_1$ and $A_2$ particles with the mass $m_1$ and $m_2$ respectively become nearly on-shell. The (\ref{StandardScale}) should be modified as
\begin{eqnarray}
    \sigma_{X \rightarrow (A_1, A_2)+Y \rightarrow Z+Y} &=& \int d \Pi_X d \Pi_Y d \Pi_Z \dots \left| \frac{i p }{ s - m_1^2 + i m_1 \Gamma_{A1} } + \frac{i q}{ s - m_2^2 + i m_2 \Gamma_{A2} } \right|^2 \dots \nonumber \\
    & \simeq & \int (d\dots) \int_{-\infty}^{\infty} ds \dots \left| \frac{i p }{ s - m_1^2 + i m_1 \Gamma_{A1} } + \frac{i q}{ s - m_2^2 + i m_2 \Gamma_{A2} } \right|^2 \dots \nonumber \\
    &=& \int (d\dots) \dots \left[ |p|^2 \frac{\pi}{m_1 \Gamma_{A1}} + 2 \text{Re} \left( p q^* \frac{2 \pi i}{m_1^2 - m_2^2 + i m_1 \Gamma_{A 1} - i m_2 \Gamma_{A 2}} \right)  + |q|^2 \frac{\pi}{m_2 \Gamma_{A2}} \right] \dots,  \nonumber \\
    & \approx & \int (d\dots) \dots \left[ |p|^2 \frac{\pi}{\bar{m} \Gamma_{A1}} + 2 \text{Re} \left( p q^* \frac{2 \pi i}{2 \bar{m} \Delta m + i \bar{m}(\Gamma_{A 1} - \Gamma_{A 2})} \right)  + |q|^2 \frac{\pi}{\bar{m} \Gamma_{A2}} \right] \dots,\label{DegenerateScale}
\end{eqnarray}
where $p$ and $q$ are the combinations of the different coupling constants involving the $A_1$ and $A_2$, respectively, and $\Gamma_{A1}$ and $\Gamma_{A2}$ are the two widths. In the last step we define $\bar{m} = \frac{m_1 + m_2}{2}$ and $\Delta m = m_1 - m_2$, so this tells us that the $\Gamma_{A 1}$, $\Gamma_{A 2}$, $\Delta m$ should be multiplied with a unified factor in order for a global rescaling upon the $\sigma_{X \rightarrow (A_1, A_2)+Y \rightarrow Z+Y}$.

Such a trick can be directly applied for the sterile neutrino creation and decay process simulations. Since in (\ref{Diag_2stage_Sterile}), the $m_{\mathcal{N}_2^{\prime}}$ and $m_{\mathcal{N}_3^{\prime}}$ are usually extremely close with each other, with the extremely small $\Gamma_{\mathcal{N}_2^{\prime}}$ and $\Gamma_{\mathcal{N}_3^{\prime}}$ to exacerbate the numerical instability problems. Define $\bar{m}_{N} = \frac{m_{\mathcal{N}_2^{\prime}} + m_{\mathcal{N}_3^{\prime}}}{2}$, and $\Delta m_N = m_{\mathcal{N}_2^{\prime}} - m_{\mathcal{N}_3^{\prime}}$, so one can zoom out $\Delta m_N$, $\Gamma_{\mathcal{N}_2^{\prime}}$ and $\Gamma_{\mathcal{N}_3^{\prime}}$ synchronously into a larger scale while keep any other parameter intact to help us conveniently compute the cross sections with the rescaling algorithm described in this subsection. We also have to note that when computing the oscillation processes (\ref{P_NDND}), (\ref{P_NDNDBar}), (\ref{P_NDBarNDBar}) and (\ref{P_NDBarND}), the $m_{\mathcal{N}_{2,3}^{\prime}}$, $\Gamma_{\mathcal{N}_{2,3}^{\prime}}$ should be recovered to their original values.

\section{Examples of the Numerical Results}

In this paper, we focus on a typical process $p p \rightarrow W^{\pm(*)} \rightarrow \mu^{\pm} N_D \rightarrow \mu^{\pm} \mu^{\pm} j j$
\cite{CMS:2018iaf, ATLAS:2019kpx, Tastet:2021vwp, CMS:2022fut}, and follow the steps described above to diagonalize (\ref{Mass_Gamma}) with the $\Gamma_{22}$, $\Gamma_{33}$ computed by the MadGraph directly, and with the $\Gamma_{23}$ acquired through the techniques in Subsec.~\ref{CrossWidthAlgorithm}. If the $\Gamma_{\mathcal{N}_2^{\prime}}$ and $\Gamma_{\mathcal{N}_3^{\prime}}$ are too small compared with the $m_{N1}$ and $m_{N2}$, techniques described in Subsec.~\ref{Rescaling} are utilized. We also have to mention that if $y_{\chi D} = y_{\chi D5} = 0$, the sterile neutrino decouples with the dark matter, therefore $\Gamma_{23} = \Gamma_{23}^{\rm SM} \ll \Gamma_{22} = \Gamma_{22}^{\rm SM} \approx \Gamma_{33} = \Gamma_{33}^{\rm SM}$ are guaranteed by the approximated lepton number conservation, where $\Gamma_{22,33,23}^{\rm SM}$ indicate the (crossed-)widths among the $\mathcal{N}_{2,3}$ in which only SM particles contribute to the decay products. Switching on $y_{\chi D}$ or $y_{\chi D 5}$ with $m_{\chi} + m_{\phi} < m_{\mathcal{N}_2^{\prime}, \mathcal{N}_3^{\prime}}$ gives rise to a significant $\Gamma_{23}$, while the (anti-)sterile neutrino might decay into the dark sector particles $\mathcal{N}^{\prime}_{2,3} \rightarrow \chi \phi$, which are assumed to penetrate all the collider facilities, behaving as the missing energies which can be the signal of the dark matter. In this paper, since we focus on the SM products, we neglect such channels.

After the event generation processes, it is fairly simple to categorize each event into the four sterile neutrino oscillation pattern described by (\ref{P_NDND}), (\ref{P_NDNDBar}), (\ref{P_NDBarNDBar}) and (\ref{P_NDBarND}). In fact, the ``.lhe.gz'' files exported by the MadGraph preserves the information of the ``mother particle'' of the final state particles, which is kept by the PYTHIA8\cite{Bierlich:2022pfr} and Delphes\cite{deFavereau:2013fsa}. One can also compute the invariant masses of the different sets of the selected final state particles, and will find out that one of the $\mu j j$ combinations with the closest invariant mass from $m_{\mathcal{N}_2^{\prime}} \approx m_{\mathcal{N}_3^{\prime}}$ originate from the same ``mother particle'', therefore the remained other $\mu$ should have been decayed from the s-channel $W^{\pm *}$. As we have verified for each of the events, these two algorithms are compatible in judging which particle originate from an intermediate nearly on-shell $\mathcal{N}_{2,3}^{\prime}$ to be called a $N$-$\mu$ in this paper, while the other muon can be called an Init-$\mu$.

The signs of the init-$\mu$ and the $N$-$\mu$ can be the hint of the initial and final state of the propagating oscillating $N_D$ and $\overline{N_D}$ system since both the vertices conserve the lepton number. If one detects a positive charged Init-$\mu$ while a positive charged $N$-$\mu$ is created, he immediately realizes that a $N_D$ was created while it had oscillated into a $\bar{N_D}$ before it decayed. In the literature, if the Init-$\mu$ and the $N$-$\mu$ are charged with opposite signs, such an event conserves the lepton number and is identified to be LNC, and on the other hand, if the two muons are charged with the same sign, such an event is recognized to be LNV.
The ratio of the cross sections between these two kinds of events is denoted $R_{ll} = \frac{\sigma_{\text{LNV}}}{\sigma_{\text{LNC}}}$.

\begin{figure}[htbp]
\centering
\includegraphics[width=0.45\textwidth]{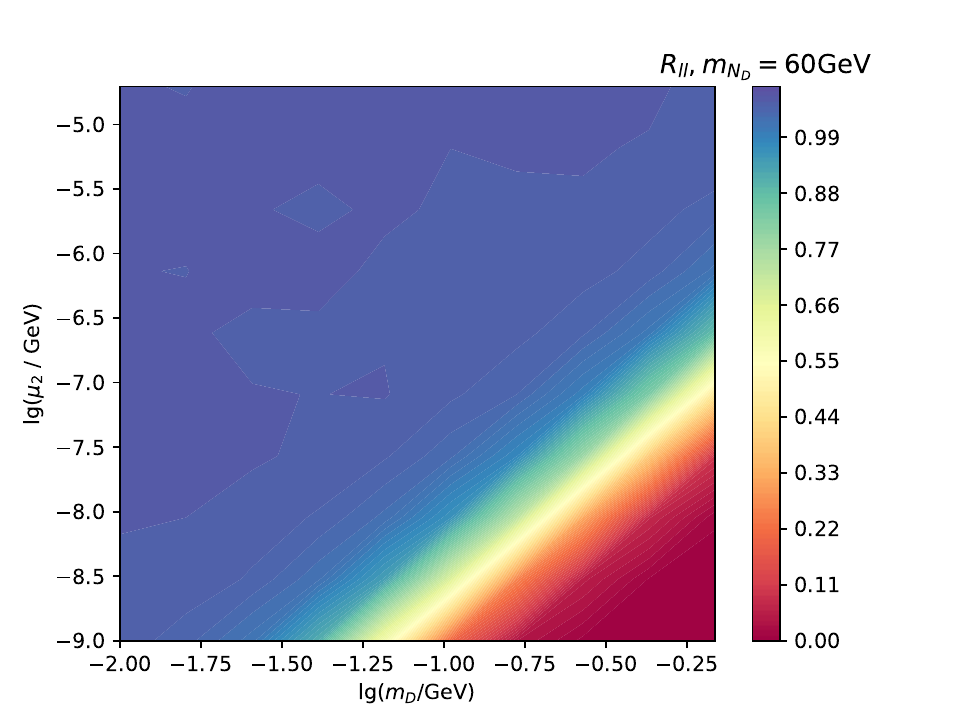}
\includegraphics[width=0.45\textwidth]{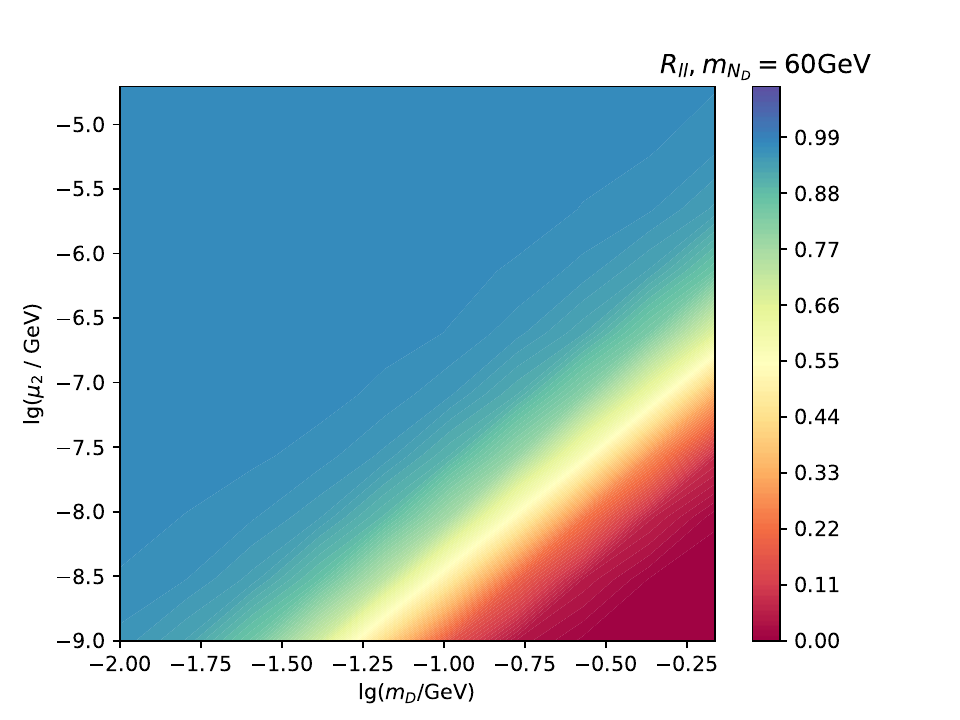}
\includegraphics[width=0.45\textwidth]{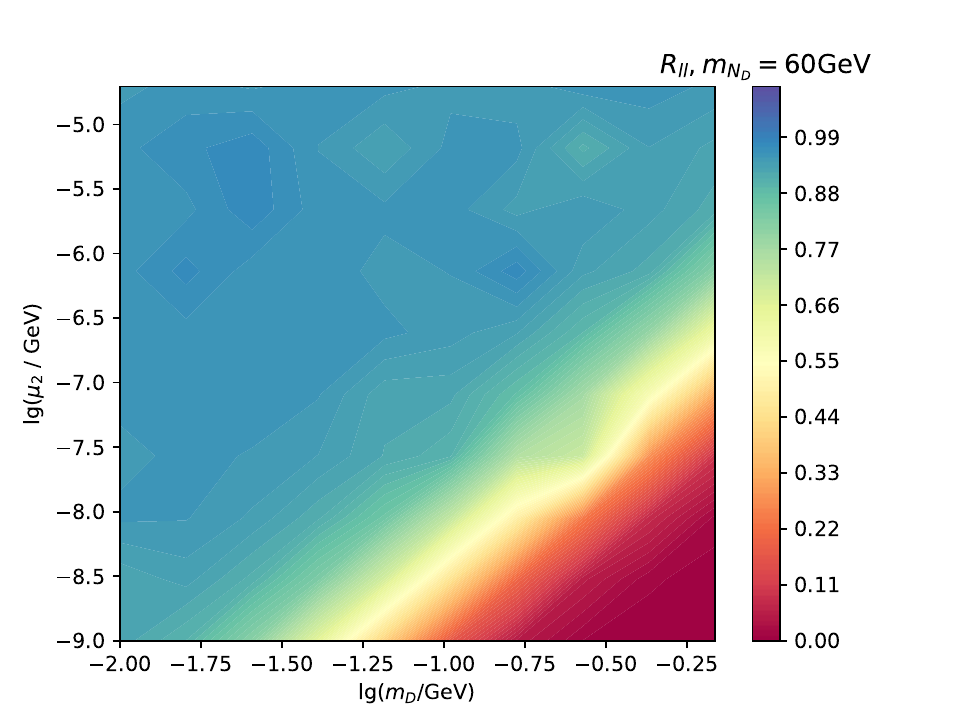}
\includegraphics[width=0.45\textwidth]{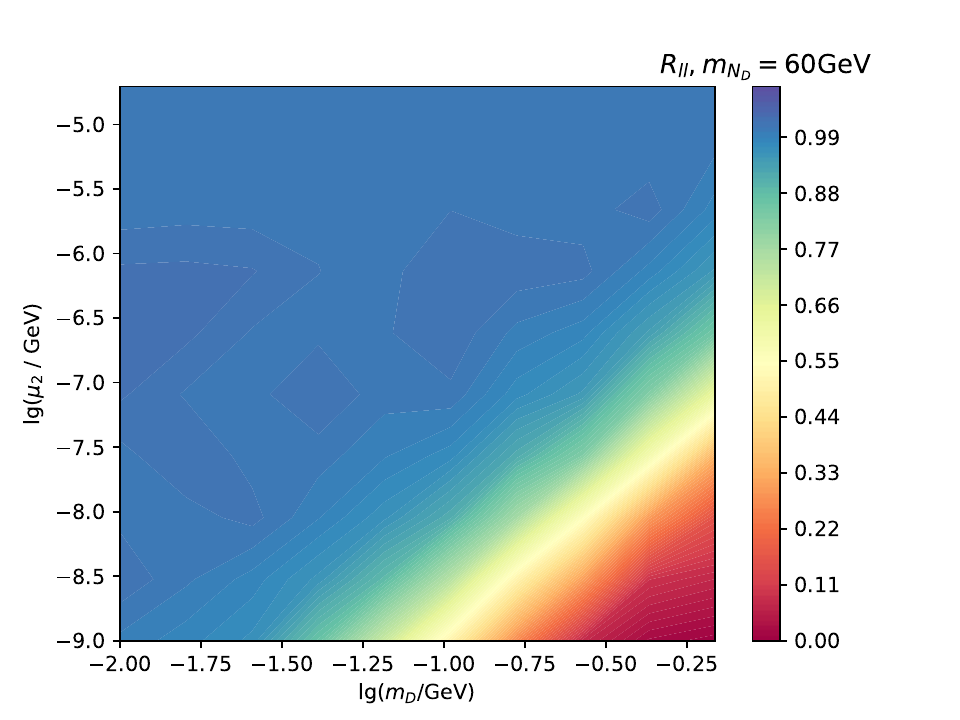}
\caption{$R_{ll}$ in the $\lg m_D$-$\lg \mu_2$ plain. $m_{N_D} = 60$GeV, and $R_{\frac{\rm DM}{\rm SM}}$ are fixed to be $0.1,~0.5,~1,~5$ in each of the four panels respectively.} \label{Rll_60GeV}
\end{figure}

\begin{figure}[htbp]
\centering
\includegraphics[width=0.45\textwidth]{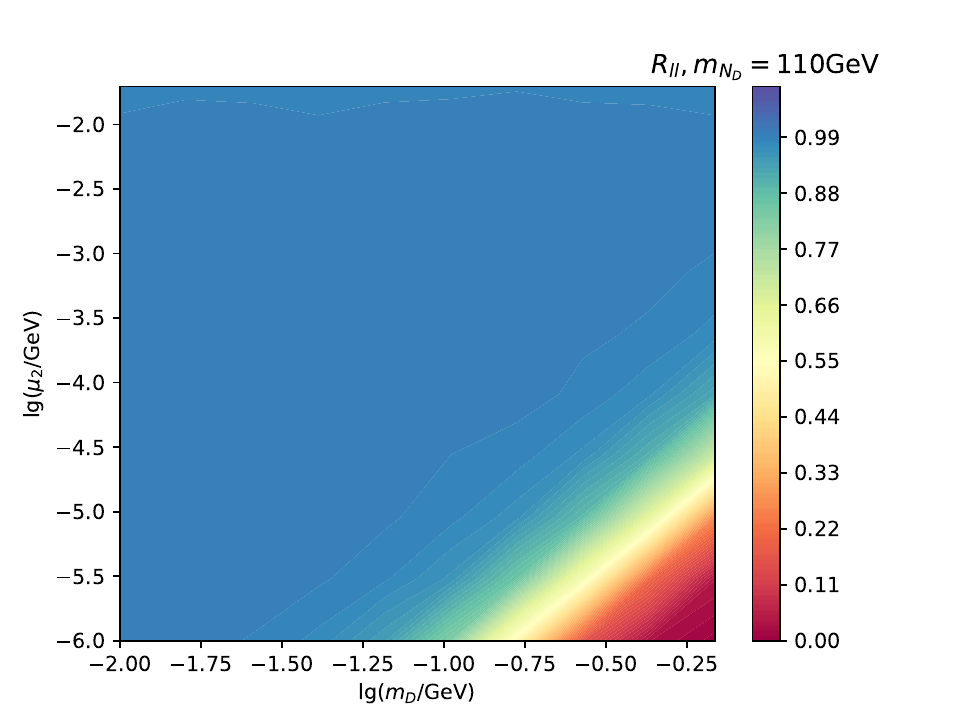}
\includegraphics[width=0.45\textwidth]{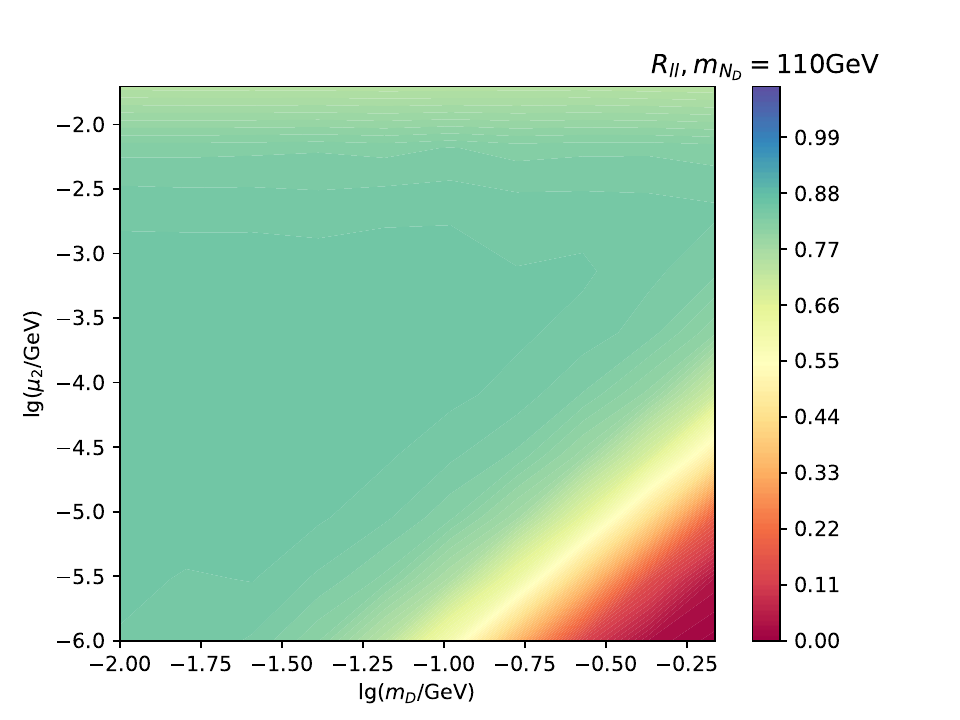}
\includegraphics[width=0.45\textwidth]{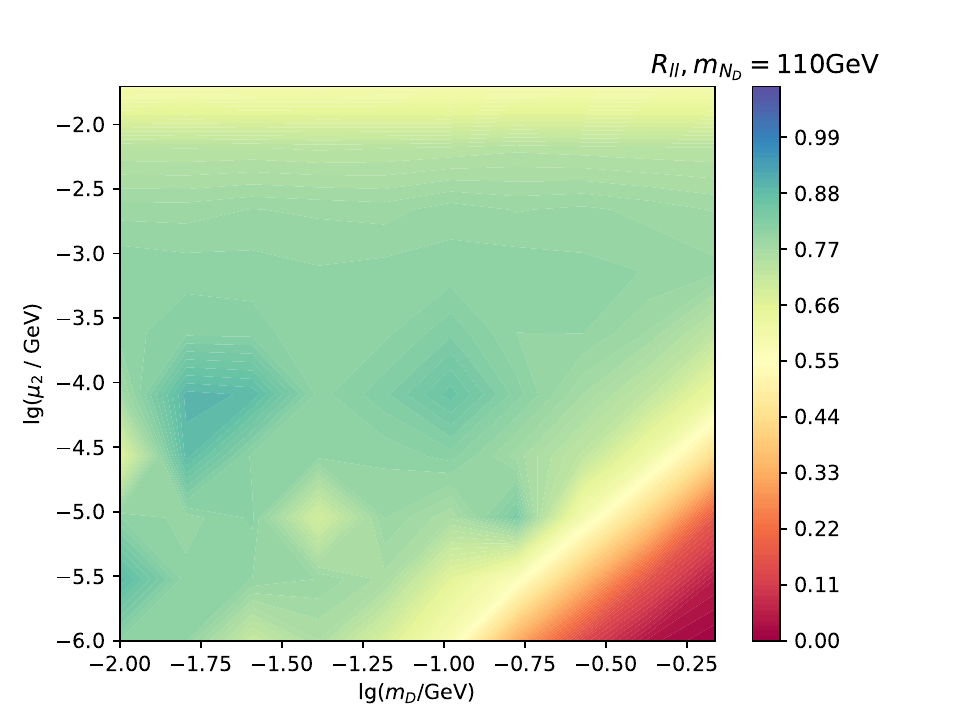}
\includegraphics[width=0.45\textwidth]{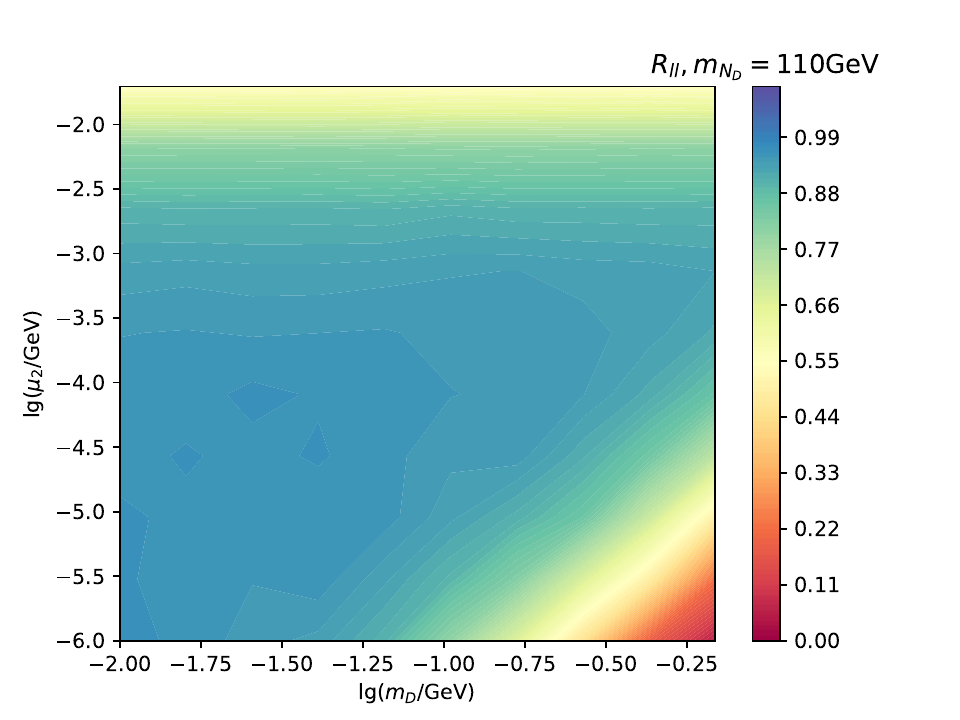}
\caption{$R_{ll}$ in the $\lg m_D$-$\lg \mu_2$ plain. $m_{N_D} = 110$GeV, and $R_{\frac{\rm DM}{\rm SM}}$ are also fixed to be $0.1,~0.5,~1,~5$ in each of the four panels respectively.} \label{Rll_110GeV}
\end{figure}

The complete parameter space involves the $m_{N_D}$, $\mu_{1,2}$, $m_{D \mu}$, $m_{\phi}$, $m_{\chi}$, $y_{\chi D}$ and $y_{\chi D 5}$, so we have to select two of them as the free parameters to present the results in a plane. Here we choose $m_{D \mu}$ and $\mu_2$ to vary while fixing the $\mu_1 = 0$ for the sake of the typical inverse seesaw model as an example, and $m_{N_D}$ is selected by $m_{N_D} = 60 \text{GeV}$ or $m_{N_D} = 110 \text{GeV}$ as the two typical values for $m_{N_D} < m_{W,Z}$ or $m_{N_D} > m_{W,Z}$, respectively. For simplicity, $y_{\chi D 5}$ is set zero. For the $y_{\chi D}$, the real part and the imaginary part are coupled with the $\mathcal{N}_2$ and $\mathcal{N}_3$ respectively, so the pure real or imaginary assignments of the $y_{\chi D}$ offers nearly disappearing crossing width $\Gamma_{23}^{\rm DM}$, where $\Gamma_{22,33,23}^{\rm DM}$ indicate the (crossed-)widths among the $\mathcal{N}_{2,3}$ contributed completely by the dark sector decay products $\phi$ and $\chi$. Since in this paper, we aim at discussing such a crossing term, so we set $y_{\chi D} \propto 1+i$ to maximize the $\Gamma_{23}^{\rm DM}$ compared with the $\Gamma_{22, 33}^{\rm DM}$, while the $|y_{\chi D}|$ is adjusted in order to fix the ratio $R_{\frac{\rm DM}{\rm SM}} = \frac{\Gamma_{23}^{\rm DM}}{\Delta M}$, where $\Delta M$ is defined to be $|\hat{m}_2 - \hat{m}_3|$. In this paper, we select $R_{\frac{\rm DM}{\rm SM}} = 0.1,~0.5,~1,~5$ for different planes in which $m_D$ and $\mu_2$ vary, and present the results of the $R_{ll}$ in Fig.~\ref{Rll_60GeV} and Fig.~\ref{Rll_110GeV} for $m_{N_D}=60$ GeV, $110$ GeV respectively. We can figure out from these panels that when $R_{\frac{\rm DM}{\rm SM}} = 0.1 \ll 1$, in quite a large range of the plain the $R_{ll}$ approaches $1$, indicating the two well-separated Majorana sterile neutrinos decaying equally into the muons with the opposite signs as a result of the completely violated lepton number. As the $R_{\frac{\rm DM}{\rm SM}}$ approaches $1$ or becomes even larger, the accumulating $\Gamma_{22,33,23}^{\text{DM}}$ ``knead'' the two resonances to overlap again, abating the $R_{ll}$ values through the interference between the resonances.

\begin{figure}[htbp]
\centering
\includegraphics[width=0.45\textwidth]{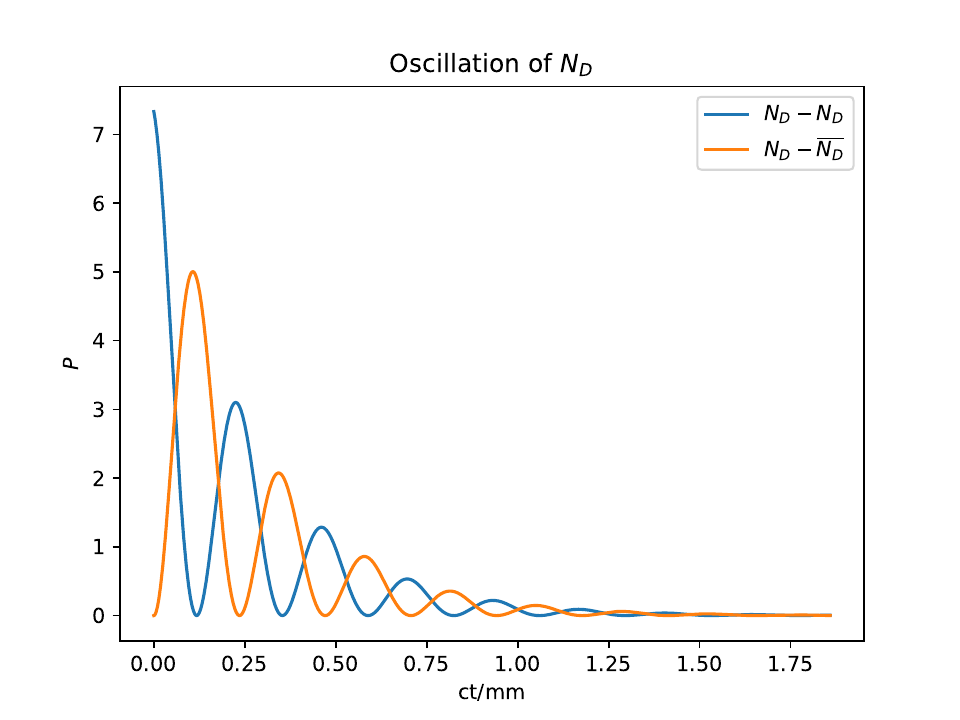}
\includegraphics[width=0.45\textwidth]{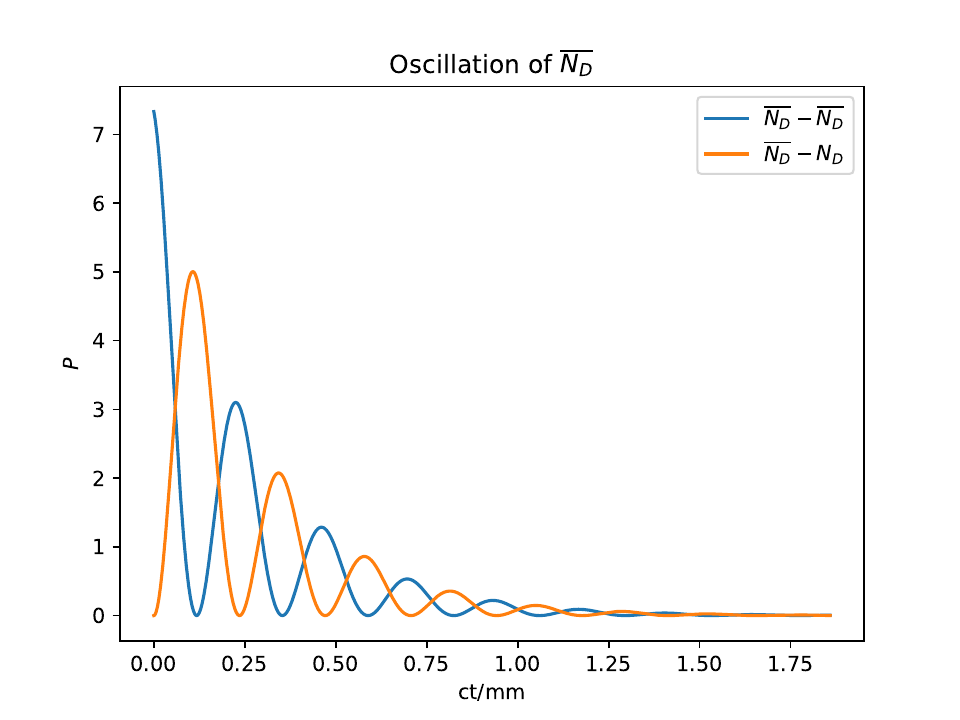}
\caption{Probabilities of the oscillations between the $N_{D}$ and $\overline{N_{D}}$ states.} \label{Oscillation}
\end{figure}

As the $m_{N_D}$ decreases below the $100$ GeV, the SM part of the width $\Gamma_{22,33}^{\text{SM}}$ drops rapidly as a result of the compression of the three-body phase space. This gives rise to the lifetime of the oscillating mediating sterile neutrino, and offers us chances to detect the displaced vertices\cite{Helo:2013esa, Izaguirre:2015pga, Deppisch:2015qwa, Gago:2015vma, Dib:2015oka, Antusch:2017hhu, Cottin:2018kmq, Cvetic:2018elt, Cottin:2018nms, Abada:2018sfh, Drewes:2019fou, Drewes:2019vjy, Dib:2019ztn, Liu:2019ayx, Cvetic:2019rms, Mason:2019okp, Das:2019fee, Chiang:2019ajm, Jones-Perez:2019plk, Lavignac:2020yld, DeVries:2020jbs, Beltran:2021hpq, Padhan:2022fak, Cottin:2022nwp, Beltran:2024twr, deVries:2024mla, Yang:2024nmk, Liu:2024fey}. Section \ref{Oscillation_Section} offers us the basic algorithm to simulate the flying time, or equivalently the flying distance of a pair of oscillating intermediate sterile neutrinos. Here we only plot a benchmark example in which  $m_{D} = 10^{-1}$GeV, $m_{N_{D}}=15$GeV, $\mu_2 = 10^{-9}$GeV, $y_{\chi D} = 5\times 10^{-6}+5\times 10^{-6} i$, $y_{\chi D5} =0 $ in Fig.~\ref{Oscillation}. There, the horizontal axis indicates the effective flying time timing the velocity of light, which can be amplified by the Lorentz factors to become the flying distance in reality if the sterile neutrino is significantly relativistic. The areas below all curves in Fig.~\ref{Oscillation} is normalized to 1, since all these curves indicate the conditional probability once we can judge the species of the initial and final states of the oscillating sterile neutrinos through the signs of the two final state muons in the .lhe files, as discussed in the Section \ref{Oscillation_Section}. With these probabilities, one thereby is able to generate the flying distance of the sterile neutrino in each of the events.

\section{Summary and Future Prospect}

Focusing on the SM products of GeV-scale the sterile neutrino, we presented our algorithm to generate the collider events of the oscillating sterile neutrino based upon an example reference model in which the crossing width among the nearly degenerate fermionic states exists. We proved the validity of our algorithm, and applied some simple tricks to utilize the original ready-made tools, such as the MadGraph. With our improvements of the oscillation formalisms in the framework of the quantum field theory, we are also able to simulate the effective flying time, or the flying distance of the intermediate sterile neutrino within an event, without modifying or patching the codes of the tools.

The algorithm and tricks described in this paper can be generalized. For an example, the sterile neutrino-portal dark matter model we rely in this paper contains some interesting parameter space in which $m_{\chi} + m_{\phi} \approx m_{N_D}$, and $m_{\chi} \approx m_{\phi}$, so that the $\chi$ and $\phi$ can co-annihilate through the s-channel sterile neutrino mediation. The Breit-Wigner resonance effects can amplify the cross sections of the annihilation processes, reducing the relic abundance of the dark matter in the freeze-out framework. Compared with the usual single Breit-Wigner resonance in the literature, in this case there are multiple nearly-degenerate s-channel mediators and oscillations among these mediators can arise, which are particularly interesting for us to study in the future.

\begin{acknowledgements}
    We thank to Yongchao Zhang, Hong-Hao Zhang for helpful discussions. YLT and QKW
are supported by NSFC under Grant No.~12005312. DYQ is supported by the National Natural Science Foundation of China (NNSFC) under Grant
No.12205387 and No.12475111.
\end{acknowledgements}

\bibliography{ref.bib}
\bibliographystyle{utphys}

\end{document}